World Scientific
www.worldscientific.com

# The Scientific Investigation of Unidentified Aerial Phenomena (UAP) Using Multimodal Ground-Based Observatories

Wesley Andrés Watters[1,2,*], Abraham Loeb[2,3], Frank Laukien[2,4], Richard Cloete[2,5], Alex Delacroix[2,6],
Sergei Dobroshinsky[2], Benjamin Horvath[2], Ezra Kelderman[2], Sarah Little[2,7], Eric Masson[2],
Andrew Mead[2,8], Mitch Randall[2,8], Forrest Schultz[2,9], Matthew Szenher[2], Foteini Vervelidou[2,10],
Abigail White[2,5], Angelique Ahlström[2], Carol Cleland[2,11], Spencer Dockal[1,2], Natasha Donahue[2,12],
Mark Elowitz[2], Carson Ezell[2], Alex Gersznowicz[2], Nicholas Gold[2], Michael G. Hercz[2], Eric Keto[2,5],
Kevin H. Knuth[2,13], Anthony Lux[2], Gary J. Melnick[2,5], Amaya Moro-Martín[2,14,15],
Javier Martin-Torres[2,16,17], Daniel Llusa Ribes[2], Paul Sail[2], Massimo Teodorani[2], John Joseph Tedesco[2],
Gerald Thomas Tedesco[2], Michelle Tu[1,2] and Maria-Paz Zorzano[2,18]

[1]Whitin Observatory, Department of Astronomy, Wellesley College, 106 Central St.
Wellesley, MA, USA 02481

[2]Galileo Project, 60 Garden Street, Cambridge, MA, USA 02138

[3]Department of Astronomy, Harvard University, 60 Garden Street, Cambridge, MA, USA 02138

[4]Department of Chemistry and Chemical Biology, Harvard University, 12 Oxford Street
Cambridge, MA, USA 02138

[5]Harvard-Smithsonian Center for Astrophysics, 60 Garden Street, Cambridge, MA, USA 02138

[6]Caltech Optical Observatory, Pasadena, CA, USA 91125

[7]Scientific Coalition for UAP Studies, Fort Myers, FL, USA 33913

[8]Ascendant AI, Boulder, CO, USA 80304

[9]Atlas Lens Co., Glendale, CA, USA 91201

[10]Department of Earth, Atmospheric, & Planetary Sciences, Massachusetts Institute of Technology
Cambridge, MA, USA 02139

[11]Department of Philosophy, Center for the Study of Origins, University of Colorado Boulder
Boulder, CO, USA 80309

[12]Athabasca University, Centre for Science, 1 University Dr., Athabasca, AB, Canada T9S 3A3

[13]Department of Physics, State University of New York at Albany
1400 Washington Avenue, Albany, NY, USA 12222

[14]Space Telescope Science Institute, 3700 San Martin Drive, Baltimore, MD, USA 21218

[15]Department of Physics & Astronomy, Johns Hopkins University, 3400 N. Charles Street
Baltimore, MD, USA 21218

[16]Department of Planetary Sciences, School of Geosciences, University of Aberdeen
Meston Building, Aberdeen, Scotland, UK AB24 3UE

[17]Instituto Andaluz de Ciencias de la Tierra (UGR-CSIC), Av. de las Palmeras
4, 18100 Armilla, Granada, Spain

[18]Centro de Astrobiología (CSIC-INTA), Madrid, Spain
*wwatters@wellesley.edu

---

*Corresponding author.

Unidentified Aerial Phenomena (UAP) have resisted explanation and have received little formal scientific attention for 75 years. A primary objective of the Galileo Project is to build an integrated software and instrumentation system designed to conduct a multimodal census of aerial phenomena and to recognize anomalies. Here we present key motivations for the study of UAP and address historical objections to this research. We describe an approach for highlighting outlier events in the high-dimensional parameter space of our census measurements. We provide a detailed roadmap for deciding measurement requirements, as well as a science traceability matrix (STM) for connecting sought-after physical parameters to observables and instrument requirements. We also discuss potential strategies for deciding where to locate instruments for development, testing, and final deployment. Our instrument package is multimodal and multispectral, consisting of (1) wide-field cameras in multiple bands for targeting and tracking of aerial objects and deriving their positions and kinematics using triangulation; (2) narrow-field instruments including cameras for characterizing morphology, spectra, polarimetry, and photometry; (3) passive multistatic arrays of antennas and receivers for radar-derived range and kinematics; (4) radio spectrum analyzers to measure radio and microwave emissions; (5) microphones for sampling acoustic emissions in the infrasonic through ultrasonic frequency bands; and (6) environmental sensors for characterizing ambient conditions (temperature, pressure, humidity, and wind velocity), as well as quasistatic electric and magnetic fields, and energetic particles. The use of multispectral instruments and multiple sensor modalities will help to ensure that artifacts are recognized and that true detections are corroborated and verifiable. Data processing pipelines are being developed that apply state-of-the-art techniques for multi-sensor data fusion, hypothesis tracking, semi-supervised classification, and outlier detection.

## 1. Introduction

Unidentified Aerial Phenomena (UAP), also known as Unidentified Flying Objects (UFOs) and Unidentified Aerospace Phenomena, have captured the public's imagination since the late 1940s but have so far attracted very little scientific interest. Since the high water mark of scientific attention in the late 1960s, the technologies required to monitor the sky and environment using every known measurement modality have been dramatically enhanced. The advent of digital cameras and inexpensive single board computers has revolutionized our ability to construct systems of sensors able to measure numerous properties of the atmosphere and of objects that pass through it. Recent advances in machine learning have produced powerful tools for characterizing "background" conditions and for detecting outlier events. In spite of this, the past 50 years have seen very few attempts by establishment scientists to design instrument systems for the express purpose of studying UAP. By contrast with most previous work on this topic, which has focused on the analysis of eyewitness reports or of observations recorded using instruments not specifically designed for scientific analysis, the present study will focus on collecting new data using calibrated instrumentation under well-understood conditions.

The term "Unidentified Flying Object," coined by Captain Edward Ruppelt who led the US Air Force investigation of UAP from 1951 to 1953, has many disadvantages that were noted long ago (Hartmann, 1972). One of these shortcomings is its inherent subjectivity ("unidentified by whom?"). It is often unclear from sighting reports that an object was necessarily flying versus floating or falling. Once identified, UFOs are commonly not "objects" at all, in the sense of materially substantial things, as in the case of light reflected from clouds, optical mirages, or radar signals reflected from density boundaries in the atmosphere (Hardy, 1972). The terms "Unidentified Aerial Phenomenon" and "Unidentified Aerospace Phenomena," which have lately come into common usage, fare only somewhat better and do not obviously encompass reports of UAP that can travel through water (Feindt, 2016). The astronomer J. Allen Hynek provided one of the most helpful definitions of this topic (Hynek, 1972b):

> "Let me define the UFO phenomenon... as that phenomenon described by reports of visual or instrumental observations of lights or objects in the sky (or near, or on the ground) whose presence, trajectories, and general character are not explainable in verifiable physical terms, even after intensive study."

A difficulty with this definition and with the acronyms UFO and UAP is that they are not descriptive: they do not refer to properties commonly ascribed to these phenomena, as if no such properties have been widely noted. "Unidentified" also suggests but does not explicitly claim that these properties are anomalous. And yet these properties are described repeatedly in the subset of witness reports that have resisted explanation even after close scrutiny, and have been attested by credible witnesses (Hynek, 1972a,b; Powell *et al.*, 2019). In such reports, UAP are commonly ascribed one or more of the following characteristics and behaviors: (i) having simple geometric shapes (spheres, discs, cubes, triangles, or cylinders) without visible wings, jets, rotors, or propellers (e.g., Condon, 1969; Sagan & Page, 1972; Powell *et al.*, 2019); (ii) remaining aloft without emitting sound (e.g., Sagan & Page, 1972; Condon, 1969), or else emitting a very high- or low-pitched sound (e.g., UKMoD, 2006; Johnson & Saunders, 2002); (iii) appearing in association with powerful magnetic fields, as suggested by aberrant compass readings (e.g., Haines, 1992; Weinstein, 2012); (iv) appearing in association with the sudden malfunction of machines and devices (e.g., Condon, 1969; Sagan & Page, 1972; Rodeghier & Longden, 1981); (v) having high luminosity with, in some cases, high-energy electromagnetic and/or particle emission, or residual radioactivity (UKMoD, 2006); (vi) demonstrating exceptional apparent maneuverability and acceleration to tens of thousands of km/h in a mere "instant" (e.g., Condon, 1969; Sagan & Page, 1972; Knuth *et al.*, 2019; Powell *et al.*, 2019); (vii) having low visibility or appearing to vanish or become cloaked (UKMoD, 2006; ODNI, 2021). Possibly the most common objection to the study of UAP is that many of these characteristics would appear at first to be impossible, and therefore must result from a misapprehension of what has actually occurred.

In spite of its disadvantages, we will use the acronym UAP throughout this paper in referring to the phenomena that our work attempts to illuminate and to which the aforementioned characteristics have been ascribed. The narrow question that we address is whether UAP partly or collectively represent phenomena currently unknown to science (a scientific anomaly). Our *wider* goal is to *determine whether there are, in or near Earth's atmosphere, measurable phenomena which can be confidently classified as scientific anomalies.* (A deeper look at what counts as a scientific anomaly is saved for Sec. 3.) Our null hypothesis is that the answer to these questions is "no," and our search will therefore be informed, in part, by theoretical and empirical scientific knowledge about known phenomena: natural phenomena, human technology, and instrument artifacts. Since we are searching for any and all "scientific anomalies" in the atmosphere, our search may uncover anomalies unlike those commonly reported as UAP. This allows us to address the widest possible range of properties that an airborne object or atmospheric or optical phenomenon may exhibit, as well as the widest range of effects that it may have on the surrounding air or local environment. Aerial phenomena with the aforementioned properties of UAP have been reported with high frequency in recent decades. At a minimum, our experiment must include the ability to capture and characterize this set of anomalous properties. The literature concerning UAP reports, in addition to our understanding of known phenomena, should therefore inform our decisions about instrument design as well as decisions about placement.

The existing literature regarding UAP is primarily concerned with UAP reports, which largely consist of eyewitness accounts. In some cases, these include direct instrument measurements of UAP, or measurements of their effects on the environment, nearby fauna, and witnesses (e.g., temporary blindness or illness). In the present work, we are not concerned with evaluating the veracity of eyewitness reports, except insofar as UAP reports can help us decide where to place instruments. Instead, our work is concerned with whether UAP can be recorded and characterized scientifically, and in this way rigorously shown to reside inside or outside of known categories. Although the conclusions of our study will not depend directly on UAP reports, our choices about instrumentation are necessarily informed by this literature. A discussion of these sources is saved for Sec. 2.3. For now, it is sufficient to note that even the use of anecdotal evidence to inform experimental design is not unusual in some scientific disciplines, such as medicine.

This paper consists of an overview of the Galileo Project's investigation of UAP. Section 2 describes the motivations for our study, which includes a discussion of recent developments in relation to the investigation of UAP, anomalies in the history of science, and the sources of information used to motivate our experimental design. Section 3 supplies a definition of scientific anomalies and how we propose to search for and recognize them. In Sec. 4 we discuss our approach to defining scientific requirements that our experiment should meet in order to achieve our objectives. The science traceability matrix (STM) is described in Sec. 5, and connects our experimental objectives to the physical parameters of objects and their corresponding observables, as well as our measurement requirements and projected instrument performance. Section 6 provides a high-level description of the entire instrument system. Considerations regarding the placement of our instruments are addressed in Sec. 7.

The use of satellite data to investigate UAP presents an additional promising avenue, and an approach similar to the one outlined in Sec. 3 (for recognizing anomalies) could be applied to this as well. A preliminary analysis of the opportunities and challenges of using satellite data is provided in a separate paper in this volume (Keto, 2023).

## 2. Motivations for the Study of UAP

In spite of numerous predictions that UAP would disappear from public attention long ago (Menzel, 1972), these phenomena have persisted and have been attested by a large number of credible witnesses (Hynek, 1966, 1972a; Cooper et al., 2019; Powell et al., 2019). This by itself may be insufficient to accept the reality of the phenomenon from a scientific perspective, but it would seem more than sufficient to motivate a systematic scientific study that is designed for the purpose. In the absence of this due diligence, it would seem premature to ignore a phenomenon to which so many intriguing properties have been ascribed. Our own study of UAP comes 75 years after reports came to wide public attention in the late 1940s. Readers may be wondering "why now?"

### 2.1. Recent history

The recent intensification of interest in UAP began in December 2017 when articles published in the New York Times (NYT) revealed allegedly credible multi-sensor observations of UAP by United States military personnel (Blumenthal, 2017; Cooper et al., 2017, 2019). These objects reportedly performed extraordinary maneuvers without a visible means of propulsion or aerodynamic control. Three videos captured by U.S. military personnel showing purported UAP were made public in 2017 and 2018 by the NYT and To The Stars (TTS), and later authorized for release by the Department of Defense in April 2020 (DoD, 2020). A NYT article also revealed a formerly secret Pentagon program established in 2007 to assess UAP as potential threats to national security, which eventually became known as the Advanced Aerospace Threat Identification Program (AATIP) (Cooper et al., 2017).

Later congressional briefings led to the establishment of a U.S. Department of Defense UAP task force (UAPTF), which released a June 25, 2021 report via the Office of the Director of National Intelligence (ODNI) containing a "preliminary assessment." This report concluded that UAP "may pose a safety of flight issue and may pose a challenge to U.S. national security," and emphasized that presently available data is insufficient to completely understand these transient and highly unusual events. This unclassified study mentioned 144 UAP reports, of which 143 remained unexplained; many of these involved corroborative observations via multiple sensors including radar, infrared, and electro-optical weapon detection systems, as well as visual observations by military personnel (ODNI, 2021).

In November 2021, the U.S. Under Secretary of Defense for Security and Intelligence established a successor to its UAP task force, called the Airborne Object Identification and Management Synchronization Group (AOIMSG) in order to "synchronize efforts across the Department and the broader U.S. government to detect, identify and attribute objects of interest in Special Use Airspace (SUA), and to assess and mitigate any associated threats to safety of flight and national security" (DoD, 2021). Legislation sponsored by a bipartisan group of U.S. senators (Kirstin Gillibrand, Rubin Gallego, and Marco Rubio) was passed by the U.S. Congress in December of 2021 directing the Secretary of Defense, in coordination with the Director of National Intelligence, to establish an office whose duties include soliciting UAP reports and mitigating the stigma that attaches to this reporting (U.S. Congress, 2021). In May 2022, the U.S. Congress held its first public hearing on the topic of UAP in over 50 years, noting that the number of reports has grown to 400 (SCCC, 2022), and that "[for] too long the stigma associated with UAPs has gotten in the way of good intelligence analysis." AOIMSG was renamed and its scope expanded to become the "All-domain Anomaly Resolution Office" (AARO) in July of 2022 (DoD, 2022a), whose purpose is to investigate "Unidentified Anomalous Phenomena" as of December 2022 (DoD, 2022b).

In August 2021 and July 2022, the American Institute of Aeronautics and Astronautics (AIAA),

the largest U.S. organization of aerospace professionals, dedicated sessions at its annual conference to the UAP topic (e.g., Graves, 2021). On June 9, 2022, NASA officials announced that a nine-month study will commence in the fall of 2022, whose purpose is to identify ways that existing data and instrument systems can be used to learn more about UAP, and to determine how new instrument systems might be designed for this purpose. The study will also consider mechanisms for funding future research (NASA, 2022).

In the past five years, major institutions connected with defense (ODNI), governance (U.S. Congress), aviation safety (AIAA), and federally-funded space science (NASA) have announced their support for the serious scientific investigation of UAP. The Galileo Project, a donor-funded research group based at Harvard University, was established in July 2021 to take up this challenge. In the following section, we consider what may be useful analogues to our present historical circumstances, which are drawn from the history of science.

## 2.2. *Anomalies in the history of science*

Philosopher and historian of science Thomas Kuhn famously argued, in the *Structure of Scientific Revolutions* (Kuhn, 1970), that most scientific research occurs within the rigid framework of a scientific "paradigm," which includes firmly-entrenched theories and dogmas (about the domain of phenomena under investigation) as well as sanctioned methods and instruments. What he termed "normal science" consists of solving "puzzles," which are well-defined problems arising in the normal course of research under an established paradigm. Puzzles that resist paradigmatic explanations can provoke what he termed "crises." According to Kuhn, one cannot even perceive a natural phenomenon as "anomalous" unless the conceptual grip of a reigning scientific paradigm is loosened during a period of crisis. Most philosophers and historians no longer accept Kuhn's account of scientific practice, with its division into normal science, crisis, and scientific revolution. Nevertheless, it is widely agreed that anomalies — phenomena that violate well-established scientific theory and dogma in unanticipated ways — are difficult to recognize for what they represent (for a discussion, see Cleland (2019), Chapter 8). For as the history of science reveals, phenomena that violate scientific expectations tend to be minimized, ignored, or even ridiculed as illusory. Yet, as the history of science also reveals, recognizing that a phenomenon is genuinely anomalous is often the first stage in the process of scientific discovery.

Danish philosopher Søren Kierkegaard famously warned that there are multiple ways to be fooled: e.g., (1) by believing in things that are not true, and (2) by refusing to believe things that *are* true (Kierkegaard, 1847). The history of science offers many examples of both follies. Kierkegaard's first scenario recalls the notorious case of Martian canals drawn as straight lines on early maps by Giovani Schiaparelli and later supposedly confirmed by Percival Lowell. After decades of frustrating attempts to replicate these observations or capture them on glass plates with telescopes of increasing sophistication, the canals were at last found to reside in the eyes of the beholder: the "canals" were optical illusions. That is, the purported anomaly was in this case resolved using a prosaic explanation, and the claims about canals on Mars were mistaken (e.g., Markley, 2005).

The history of science also offers many examples of the earlier-mentioned second scenario from Kierkegaard, where epistemic authorities refuse to acknowledge phenomena for which compelling evidence existed a long time, but which reside too far outside the reach of their explanatory powers. An important example in light of present controversies surrounding UAP is the response of Enlightenment philosophers to reports that rocks occasionally fall from the sky and have a celestial origin. Despite a long historical record of these events, meteorite falls were largely denied or else attributed to lightning strikes ("thunderstones"), spontaneous accretion of stones in the atmosphere, or debris ejected from distant terrestrial or lunar volcanoes (Marvin, 2006). Renowned chemist Antoine-Laurent de Lavoisier famously signed a report to the Royal Academy of Sciences in Paris, which concluded that it is impossible for stones to fall from the heavens (Fougeroux *et al.*, 1777). Some natural philosophers belittled eyewitnesses as superstitious and many accounts of meteorite falls were dismissed as folk tales (Marvin, 2006; Gounelle, 2006; Marvin, 2007). These views finally softened after a series of well-documented sightings in the late 18th century and the 1803 mass sighting of a spectacular fall over L'Aigle in France, reported by a large number of witnesses deemed credible and interviewed by physicist Jean-Baptiste Biot (Biot, 1802; Gounelle, 2006). The discovery of asteroids in the first decade of the 19th century provided a source for meteoritic materials, and compositional evidence eventually convinced natural philosophers of a cosmic origin (Marvin, 2007).

To which of these categories do UAP belong? That is, are they true anomalies that suggest modest or profound changes are needed to the way scientists think about the world? Or are they Kuhnian "puzzles" or mere illusions that can be explained in terms of current scientific beliefs, deriving mainly from accidents of mistaken identity, instrument artifacts, wishful thinking, fraud, or some combination of these? In light of evidence from the past 75 years, our investigation begins without a *scientific* answer to this question. On the other hand, the events of the past five years described above, along with seven decades of remarkable reports that have resisted satisfactory explanation, as well as a history of intellectual revolutions that were preceded by stubborn refusals to examine inconvenient evidence, would at least suggest that UAP should be studied scientifically.

In 1969, the American Association for the Advancement of Science (AAAS) convened a symposium of distinguished scientists in Boston, MA, to decide whether the UAP problem deserved scientific attention. The substance of the debate was captured in a book edited by Carl Sagan and Thorton Page (Sagan & Page, 1972). In many important respects, the debate has not substantially changed since this book was published, in spite of confident predictions that the entire phenomenon would disappear (Menzel, 1972). In Appendix A, we consider the objections raised by some of the participating scientists, as well as common objections and misconceptions still prevalent today.

## 2.3. *Sources of information*

There is an abundance of anecdotal reporting with regard to UAP and only a few instrument-based studies. In the USA, investigations by the US Air Force into UAP began with Project Sign in 1948, in response to public reports from pilots and others, with Project Grudge following in 1949. The most significant government study was Project Blue Book (1952–1969), which investigated over 12,000 military and civilian reports of UAP (USAF, 1995). An inquiry addressing the adequacy of Project Blue Book was funded by the Air Force and conducted by a group at the University of Colorado, with the resulting Condon Report concluding that further study of UAP would be of marginal scientific utility (Condon, 1969). Both Project Blue Book and the Condon Report included publicly available descriptions of UAP sightings, along with suggestions about how to perform useful measurements to characterize these phenomena.

Many of the presentations and proceedings from the 1969 AAAS symposium focused on discussion of cases from Project Blue Book and the Condon Report (Sagan & Page, 1972) and a range of views were aired. A presentation by atmospheric scientist and National Academy of Sciences member James McDonald titled "Science in Default" noted that about one-third of the cases examined by the Condon Report were classified as "unexplained." McDonald analyzed four such cases, concluding that the methods used by previous Air Force studies and the Condon committee failed to meet standards of scientific rigor (McDonald, 1972). Along with astronomer J. Allen Hynek, who served as consultant for the Air Force investigations, McDonald is one of the few academic scholars to have conducted independent scientific research addressing UAP reports. A later scientific panel was convened in 1997 and consisted of UAP researchers and mainstream scientists. Headed by Stanford University physicist Peter Sturrock, this group concluded that the UAP phenomenon was a complex mystery warranting serious scientific attention, which should be focused on the examination of physical evidence (Sturrock *et al.*, 1998).

In addition to government reports from USA-based agencies, government agencies in the UK (Project Condign: Unidentified Aerial Phenomena in the UK Air Defence Region (UKMoD, 2006)), France (GEIPAN: Groupe d'Études et d'Informations sur les Phénomènes Aérospatiaux Non-identifiés, or Unidentified Aerospace Phenomenon Research and Information Group (GEIPAN, 2022)), and Canada (Project Magnet (Smith, 1968)), have also produced reports acknowledging the existence of UAP as an unsolved problem. Finally, some government reports have alluded to classified observations which cannot be evaluated or addressed by the scientific community (e.g., ODNI, 2021).

The published, peer-reviewed scientific literature includes only a handful of studies that have attempted to identify or characterize anomalous aerial phenomena. Some of these studies have drawn attention to observations of unusual atmospheric phenomena and have suggested plausible natural explanations (e.g., Zou, 1995; Pettigrew, 2003). A few studies have described observations of potentially anomalous phenomena acquired using conventional instrumentation, while also offering explanations that reside completely inside the framework of conventional atmospheric science and astronomy (e.g., Cen *et al.*, 2014; Villarroel *et al.*, 2021). Other refereed studies have analyzed data from previous UAP reports and have published their results in academic journals (e.g., Maccabee (1994, 1999), Vallee (1998), Knuth *et al.* (2019)). These have described or synthesized previous observations, but all have concluded that insufficient data exists to explain at least some reports or predict the likelihood of future sightings of UAP. Only a few peer-reviewed studies have described integrated instrumentation for the systematic study of UAP (Stride, 2001), instrumented field investigations of UAP (e.g., Teodorani, 2004), and forensic investigations of UAP-related remnant materials (Nolan *et al.*, 2022).

Three well-documented, long-term scientific field studies are worth noting. (1) Hessdalen Valley, Norway, 1984-present (Strand, 1984; Teodorani & Strand, 2001; Teodorani, 2004; Hauge, 2007): spanning over two decades, distinct teams of researchers captured radio emissions, magnetic field anomalies, optical video and still images, and low-resolution optical spectra ($\lambda/\delta\lambda < 1000$) of episodic anomalous atmospheric lights. Additional instrumentation included active radar, a seismograph, and particle counter. (2) Yakima Valley, WA, 1972–2007 (Akers, 1972, 1974, 2001, 2007): an episodic, multi-year investigation of anomalous light phenomena, which focused on credible local eyewitness reports and concomitant instrument measurements. A total of 82 events were documented, along with supportive instrument recordings: eight 35 mm color slides, one analog video with sound, and over 12 h of digital magnetic data. Instruments included radio standard-time and quartz regulated audio time-base references, cassette audio recorders, three visible/near-infrared/spectrogram-capable 35 mm cameras, a 16 mm motion picture camera, one analog video/audio camcorder, two analog and digital logging magnetometers, hand compasses, Geiger counter, wide-coverage radio monitor receiver, experimental near-infrared audio photometer, compass spin detector, and ultrasound-to-audible translating microphone. (3) Piedmont, Missouri, 1973–1981 (Rutledge, 1981): a study of anomalous atmospheric light phenomena using binoculars, cameras, camera-equipped astronomical telescopes, radar transceivers, a spectrum analyzer, and spectroscopic camera. This work yielded 153 detailed observer sightings and 33 photographs of UAP. Two additional studies from the early 1950s, Project Twinkle (USA) and Project Magnet (Canada), are described in Ailleris (2011).

The majority of the aforementioned field studies, like the majority of written information regarding UAP, is contained in a vast "gray literature" that includes books, conference papers, technical reports, websites, and unpublished materials. This gray literature forms the bulk of available information beyond declassified military and government reports. In addition to non-refereed documentation of instrument-based scientific field studies of UAP like those described above, this includes compilations and meta-analyses of UAP reports (e.g., Poher & Vallee, 1975; Vallee, 2007) and proposals defining standards and procedures for methodical research (Vallee, 2014). This literature is also a rich source of information about measurements that could have been used to characterize and potentially identify the phenomena described in UAP reports, and hence can usefully inform the design of instrumentation for the study of UAP. The quality of studies from the gray literature varies greatly, with many sources falling short of the standards of rigor or peer review that are characteristic of academic science. This leads to an important question regarding the quantification of uncertainty: if we do not know the quality of information contained in a given source, then how should we decide whether to consider or rely on the source?

The design of an experiment intended to search for anomalies in the Earth's atmosphere will require specific choices regarding the sensors and methods to be used, which in principle should be driven by previous reports of anomalous observations in addition to observations of known phenomena. In practice, the gray literature in any field tends to be ignored by academic researchers because this often does not meet the evidentiary standards required for academic science. At the same time, there is no epistemological reason that reports outside of the peer-reviewed literature could not contain information useful for the design of a scientific experiment. For example, publicly available reports are a potentially useful source of anecdotal evidence, as in medical research when anecdotal evidence is used to inform rigorous experimental design (Campo, 2006).

As the present work seeks to establish a firm foundation for scientific research in a field with few prior academic publications, a smattering of helpful declassified reports, and an enormous gray literature, it must draw on several sources of information to inform experimental design: (1) peer-reviewed academic literature about known aerial objects and phenomena (human technology, natural phenomena); (2) the few academic peer-reviewed studies regarding UAP; (3) government agency and government-commissioned analyses of UAP; and occasionally (4) sources from the gray literature describing aspects of UAP not otherwise addressed by sources in the foregoing categories. It should be emphasized that the conclusions of our work will not depend on any previous claims that we cannot independently confirm for ourselves. That is, we will not take for granted claims from any of the sources mentioned above. Instead, we shall use some of these references to inform choices about instrumentation, in order to increase the chances of capturing previously-described phenomena that are allegedly anomalous.

The Galileo Project's UAP investigation has taken very seriously one of the most important lessons from all of the preceding literature: that it is crucial to make simultaneous and corroborative observations using a multimodal, multispectral suite of calibrated instruments that are completely understood and entirely under the control of scientific investigators. This will ensure the highest likelihood of recognizing instrument artifacts and of enabling independent experiments to reproduce our measurements. Finally, it is worth emphasizing that this work will not address many attested aspects of UAP sightings and encounters, such as long-term physical effects on the environment and nearby fauna, or effects on the health, psychology, and awareness of witnesses (e.g., Vallee, 1975; Vallee & Davis, 2004; Vallee, 2007).

## 2.4. Contemporary research efforts

We are aware of three contemporary, ongoing scientific efforts to detect and characterize UAP. (1) The Interdisciplinary Research Centre for Extraterrestrial Studies at Julius–Maximilians–Universität Würzburg, Germany, is developing a system of automated, continuously operating optical and infrared cameras, with machine learning

(ML) software designed to detect UAP (Vodopivec & Kayal, 2018; Kayal, 2020). (2) UAPx is a group of scientists, technicians, and ex-military UAP witnesses, whose instrumentation includes optical and thermal infrared (IR) cameras and high-energy particle detectors. UAPx conducted a week-long field investigation in the vicinity of Catalina Island off the coast of Los Angeles in 2021 (Knuth & Szydagis, 2022). (3) The UAP technical commission of the Association Aeronautique et Astronautique de France, also known as "Sigma2," makes use of existing aerial detection networks, including active and passive bistatic radars, visible and thermal infrared cameras, optical and radio meteoroid surveillance networks, IR and optical imaging satellites, as well as gravimeter and magnetometer networks (SIMGA2, 2021). As of this writing, none of these ongoing efforts have reported UAP detections in peer-reviewed scientific literature.

Meteor observatories and their networks have developed technology that is highly relevant to the design of UAP observatories (e.g., Vida et al., 2021). Of special note is the Fireball Recovery and Inter-Planetary Observation Network (Colas et al., 2020), which uses cameras and bistatic radar to detect aerial objects in the airspace of France and other participating countries, and whose data is used by the Sigma2 commission (SIMGA2, 2021). This and five additional large-scale fireball and meteor tracking networks are currently in use around the world; the details of their operational characteristics are described in Szenher (2023), in this volume.

## 2.5. The Extraterrestrial Hypothesis

In the past several decades, communities of scientists including astrobiologists, planetary scientists, and astronomers have been searching for signs of extraterrestrial life and extraterrestrial civilizations (ETC). So far, these efforts have focused primarily on the search for potentially habitable environments in the solar system (e.g., Shapiro & Schulze-Makuch, 2009; McKay et al., 2014) and extrasolar planets (e.g., Schwieterman et al., 2018; Meadows et al., 2018) as well as radio and optical signals from remote ETCs (e.g., Cocconi & Morrison, 1959; Howard & Horowitz, 2001). This interest has strengthened in recent years following discoveries by the Kepler spacecraft that potentially habitable planets are superabundant in our galaxy (Lissauer et al., 2014). According to some estimates, a well-resourced ETC could send probes to explore the entire galaxy in a fraction of its 10 billion year lifespan, using methods that are within reach of near-future human technology (see Lingam & Loeb, 2021, for a review). Studies in the peer-reviewed literature have evaluated the prospect of finding evidence for ETC artifacts in our solar system (e.g., Haqq-Misra & Kopparapu, 2012; Kecskes, 2013), including a search for evidence of artifacts in terrestrial orbit before the advent of human satellites (Villarroel *et al.*, 2022).

UAP are almost automatically associated in the public imagination with an extraterrestrial origin. If the properties and behaviors that have been ascribed to UAP are eventually verified, then considering that these properties and behaviors in some cases dramatically exceed the performance capabilities of known human technology, it is not unreasonable to wonder if UAP could represent examples of extraterrestrial technology. Although the Galileo Project was established to search for extraterrestrial artifacts in the solar system, and although UAP are the target of the present investigation, we begin this research with no assumptions or expectations about the nature and provenance of UAP. The goal of the Galileo Project's UAP investigation is initially broader in scope and more foundational: it is to determine whether there are measurable phenomena in or near Earth's atmosphere which can be confidently classified as *scientific anomalies*. We turn next to defining this term.

## 3. Recognizing Scientific Anomalies

The primary science goal of the Galileo Project's UAP investigation is *to determine whether there are measurable phenomena in or near Earth's atmosphere that can be confidently classified as scientific anomalies*. This section will define "scientific anomaly" as well as describe how a scientific anomaly can be recognized, and how this is distinguished from philosophical and colloquial uses of "anomaly."

Thomas Kuhn was the first philosopher of science to place anomalies at the center of the process of scientific discovery (Kuhn, 1970). According to Kuhn, an anomaly is a phenomenon that violates the expectations of a scientific paradigm, leading in some cases to the replacement of a scientific theory. In contrast to Kuhn, we define a "scientific anomaly" as a phenomenon that is unknown to current science and that persists in resisting explanation in terms of widely accepted scientific beliefs.

Such beliefs are not necessarily consequences of fundamental science — they could be scientific *dogmas*: widely accepted philosophical assumptions or empirically-derived generalizations that are not founded in theory. An example of a scientific dogma was the widespread 18th century belief that there are no planets beyond Saturn, even though this is not precluded by Newtonian theory. The upshot is that a "scientific anomaly" (in our sense of the word) need not lead to a major change in a scientific theory, such as the standard model of physics. Moreover, a scientific anomaly in our sense is at first recognized statistically: its properties lie outside of the quantified phenomenological envelope of all known phenomena, and appear to arise from a previously-unknown generative process. Detecting statistical outliers is a step on the way to recognizing scientific anomalies. The latter results if a statistical anomaly resists explanation in terms of prevailing scientific beliefs.

The remainder of this section is devoted to outlining a procedure for recognizing scientific anomalies. To begin, we require for this task a few additional definitions and clarifications. The word *object* will be used in the sense of a thing that has been observed and is apparently localized in space; the words *phenomenon* and *object* will be used interchangeably throughout the remainder of this text. *Event* is used to refer to the period during which an object was detected by instrumentation. Events are triggered by a *detection*: an observation that stands out from background measurements (Cloete, 2023, this volume), such as a distinctively-shaped fast-moving object in an all-sky camera frame. The present study will focus on events that are concurrent with, or can be ascribed directly to, the appearance of aerial objects in camera and radar data.

The quantities measured during an event are called *observables* and the recorded data are called *event data*. *Physical parameters* are the intrinsic properties of objects or of their local interactions with the atmosphere, which are derived from measurements of the observables comprising event data. For example, the local magnetic field strength is an observable. When combined with range estimates, it is possible to estimate the object's magnetic dipole moment, which is a physical parameter. Apparent brightness is an observable that may vary throughout an event; combined with range information, this can be used to infer time-dependent luminosity. Finally, it should be noted that some observables may be associated with objects in the sky, while others may be associated with the environment, or with the effects of an aerial object on the environment, or all three of these.

We will use the word *features* to refer to all of the quantities that can be used to distinguish one object or event from another: i.e., observables, physical parameters, as well as quantifiable characteristics that are explicitly or implicitly defined by statistical models and machine learning models. That is, in some cases features will be relatively straightforward to define and understand, such as maximum acceleration, luminosity, and size. Others may be defined implicitly by a model. For example, deep neural networks (DNNs) can be trained to generate features that are highly effective for distinguishing objects in photographic imagery (Rawat & Wang, 2017) or acoustic signals in time-dependent power spectra (Xie *et al.*, 2016; Kulyukin *et al.*, 2018). The meanings of features that are implicitly defined in this way may be difficult to interpret. It should be emphasized that feature selection is an iterative process: while our data are being collected, we will revise the set of features that are most effective for distinguishing categories of phenomena. The set of feature values associated with an event or an object observed during an event is called a *feature vector*.

The following describes the key steps in our search for scientific anomalies. (For a more detailed discussion of algorithms, see Cloete (2023) in this volume.) This approach is designed to answer what might be called "Menzel's Challenge" (Menzel, 1972) which can be paraphrased as follows: *What phenomenon did the UAP most closely resemble?* and *Why do you believe that the UAP was not this phenomenon?* The steps are:

(1) *Perform a census of aerial phenomena:* Spanning years and multiple locations, our instruments will gather data regarding phenomena that reside within or which pass through the atmosphere. Feature vectors associated with those objects and events will be measured and recorded in a database. In order to be as comprehensive as possible, the census should be conducted in a diverse set of locations and under a wide range of conditions. Important considerations for selecting instrument sites from which to conduct the census are discussed in Sec. 7.

An example of a limited census of conventional aircraft is shown in Fig. 1. These data derive from a four-day sample of flight characteristics of conventional aircraft recorded from Automatic Dependent Surveillance-Broadcast (ADS-B) transponder messages. Only two features are plotted: ground speed and altitude. After this short interval, overdense regions corresponding to aircraft types and flight modes have become apparent, and the outline of a performance envelope (or "phenomenological envelope") has begun to take shape.

(2) *Validate the census data:* A continuous and systematic review of the event data is required to validate the detections before they are added to the census database. This is important because our intention is to publish data that will not require instrumentation expertise for analysis and evaluation. Data validation involves the identification of detections that should not be included in the census database (and instead ignored or logged in a separate database) because they (i) do not correspond to phenomena in the sky or environment (e.g., instrument artifacts), or (ii) were not characterized with sufficient accuracy or resolution to be useful. Feature vectors in the first category correspond to instrument artifacts, malfunctions, and errors. Examples in imagery include dust doughnuts, hot pixels, lens flares, cosmic ray strikes, and thermal noise. Simple detection algorithms and more sophisticated supervised machine learning models can be used to recognize such detections and mark them as artifacts, or simply ignore them when detections are recorded in the first place. Feature vectors in the second category (low accuracy and/or resolution) are labeled "under-determined." This will happen in many cases because: (a) the uncertainties of measured features are too large (e.g., an acceleration with large bounds, or a poorly-resolved image of a very distant object) or (b) some key features were not measured (e.g., the detection was insufficiently corroborated by other instruments). It is vital to draw a distinction in this way between genuine outliers and measurements that were merely ambiguous because they were poorly resolved.

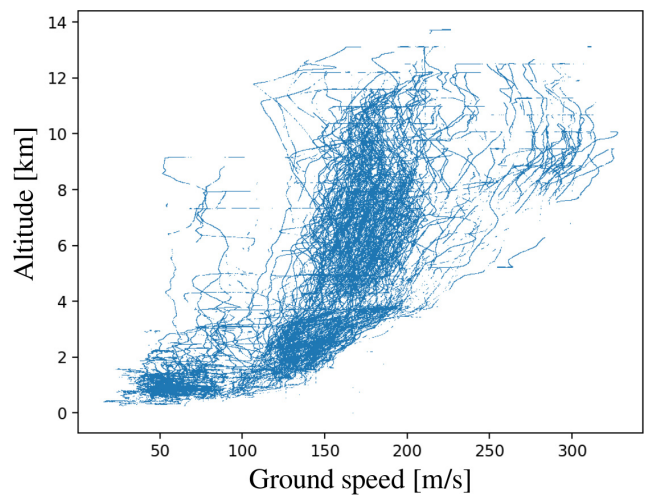

Fig. 1. Census of ∼ 1200 aircraft whose ground speed and altitude were recorded during a four-day interval in November 2021 over Wellesley, MA, USA. Even this short period and small sample size is sufficient to show the overdense cores of clusters and the emerging boundaries of a performance envelope.

Table 1. Categories of known phenomena with examples. Expected detection modalities in addition to optical and thermal infrared imagery: [*]acoustic emission; [†]radio emission; [‡]radar echoes.

| Category | Subcategory | Examples |
|---|---|---|
| naturally-occurring | fauna | birds,[*] bats,[*] insects[*] (esp. fireflies); flocks[*‡] and swarms[*‡] |
| | atmospheric phenomena | cumulus clouds,[‡] lenticular clouds,[‡] fallstreak holes (a.k.a. hole-punch clouds), lightning,[*†] sprites,[*†] ball lightning,[*†] St. Elmo's Fire,[*†] aurorae[*†‡] |
| | atmospheric optical phenomena | parhelia (sundogs), rainbows, mirages, reflected surface lights |
| | astronomical phenomena | planets, the Moon, stars, comets, asteroids, meteors,[*‡] bolides[*‡] |
| human-made | powered piloted vehicles[*†‡] | jet airliners, multi-engine propeller aircraft, single-engine propeller aircraft, helicopters, military and stealth aircraft |
| | drones[*†‡] | copter drones, fixed-wing drones |
| | lighter-than-air | hot air balloons,[‡] blimps,[*‡†] party balloons, lanterns |
| | other unpowered flight | gliders,[‡] paragliders,[‡] parachutes,[‡] kites |
| | light displays | fireworks,[*] flares, laser beams, aerial vehicle signal and warning lights, automotive vehicle lights on raised roadways, lighthouses, boats,[*†‡] oil rigs,[†‡] flood lights |
| | artificial satellites[†‡] | satellite trains, space stations, space junk, reentering space debris |
| | rockets[*‡] | missiles (weapon systems), rockets (for spaceflight) |
| | exhaust | jet contrails |
| | waste | plastic bags, paper fragments |

(3) *Categorize phenomena in the census:* Feature vectors span a high-dimensional space in which phenomena of specific types are expected to form clusters. Judicious selection of features for cluster analysis is vital, as some features will not be especially useful for discriminating categories of phenomena. Alternatively, well-worn techniques such as Principal Component Analysis (PCA) can be used to reduce the dimensionality of the feature space before proceeding. Clusters can be defined using an array of well-established approaches based on the proximity or density of feature vectors, such as Random Forests (Ho, 1995), k-Means (Lloyd, 1982), and k-Nearest Neighbors (kNN) (Fix & Hodges, 1989). Assigning categories to clusters can be accomplished using semi-supervised learning, through which a small set of human-defined labels are attached to entire clusters by association (e.g., "airplane," "helicopter," "bird"). Classification can also benefit from ensemble learning and combining results of multiple inference models. The clusters themselves will be associated in turn with the most important umbrella categories of aerial phenomena (see Table 1): (a) naturally-occurring phenomena (e.g., clouds, birds); (b) human-made phenomena (e.g., airplanes, helicopters, lofted litter); and (c) other. Clusters in the "other" category cannot be confidently assigned to either of the two preceding categories, and warrant further attention as potential corroborated outliers (see below).

(4) *Identify statistical outliers:* As the census is populated, we may discover that a few feature vectors cannot be identified as belonging to any known category because they are statistically distinctive. These are termed *outliers*. Outlier detection has been the focus of intensive research for the past two decades with applications to many problems including fraud and fault detection (Singh & Upadhyaya, 2012). Outlier detection is widely acknowledged to be a challenging problem for many reasons. In the present case, we do not know the generative distribution function for outliers, and we will not even know the generative distribution function for nominal values in the case of many features. Many of our features are also likely to exhibit noise, which cannot be easily distinguished from outliers, especially if the generative nominal distribution function has long tails. Moreover, some outliers may be heavily context-dependent: e.g., some observations could be normal in winter or daylight and abnormal in summer or nighttime.

The approaches to outlier detection can been grouped as follows (Emmott *et al.*, 2015): projection-based methods (e.g., Isolation Forests (Liu *et al.*, 2008)), neighbor-based methods (e.g., kNN Angle-Based Outlier Detection (Kriegel *et al.*, 2008)), density-based methods (e.g., Robust Kernel Density Estimation (Kim & Scott, 2012)) and model- or quantile-based methods (e.g., Support Vector Data Description (Tax & Duin, 2004)). Most approaches to outlier detection involve computing an outlier score for each feature vector, which is larger for a higher outlier tendency. Outlier scores will be calculated using multiple approaches (i.e., model ensembles), using subsets of the census data, and subsets of features. Simulation of event scenarios using synthetic data may be required to identify approaches that are more likely to be effective and robust. Moreover, the uncertainty ellipsoid surrounding feature vectors, when available, must be consulted as part of the outlier detection algorithm, to ensure that highly uncertain measurements are not confused with being special in some way.

An outlier detected with only one measurement modality cannot on its own constitute a scientific anomaly or anomaly candidate. Feature vectors that are outliers in a single dimension but which represent observations involving multiple measurement modalities (e.g., optical and acoustic) are called *verified outliers*. Feature vectors that exhibit multiple outlier values (i.e., which have high outlier scores along multiple dimensions) are of special interest because their unusual character is apparent in multiple ways; these are called *n-fold outliers*, where $n$ is the number of individual outlier features. Verified $n$-fold outliers with large $n$ are naturally of special interest.

(5) *Identify corroborated outliers:* Long-term census observations may reveal that outliers eventually gain proximate neighbors from distinct events which together form new clusters. Such a cluster of unclassified feature vectors may represent repeated observations of the same or similar phenomenon. Clusters of feature vectors that remain unidentified are called *corroborated outliers* or *novel clusters* since they may cease to be outliers according to some statistical criteria (e.g., nearest-neighbor distance).

(6) *Reject outliers using known performance envelopes:* In spite of encompassing an enormous database of feature vectors, the census may not represent all of the most extreme characteristics of any category of known phenomena. For this reason, a look-up table or hypersurface of performance characteristics for extreme cases of known phenomena will be used to reject corroborated outliers that reside entirely within these previously-characterized regions of the feature space. For example, an object may exhibit an acceleration exceeding the maximum linear acceleration of any vehicle in the census, but fail to exceed that of the most advanced missile technology, in which case it may be rejected. On the other hand, if it also displays capabilities outside the performance envelope of said missile (e.g., the ability to hover, or to turn with a smaller turn radius), then it is maintained. Corroborated outliers that survive this rejection step are designated candidate statistical anomalies (or "candidate anomalies," for short). Unfortunately, it may be impossible in many cases to find published descriptions of the high-dimensional performance envelope for specific technologies; for this reason, a detailed census may be our best guide in most situations. An additional challenge is that novel technology is commonly developed in secret and may only slightly exceed performance limits that are publicly known; candidate anomalies must therefore dramatically exceed these limits in order to be highly convincing.

(7) *Perform targeted follow-up observations:* It may happen that candidate anomalies exhibit characteristics that can be studied in greater detail with modified instrumentation or by adding new instrumentation. This can be postponed for a future phase of the investigation but also may, depending on the level of challenge, be incorporated during the census phase. For example, if corroborated outliers exhibit a specific radio frequency emission, then follow-up observations may interrogate this spectral band with higher frequency resolution. This could make possible the detailed recording of characteristic behaviors and properties, allowing even more precise features to be defined. This may lead over time to a relatively detailed characterization of the candidate anomaly and predictions regarding the conditions in which they are most likely to appear.

(8) *Share results for independent corroboration:* If the candidate is repeatedly observed and carefully characterized (in some cases with targeted follow-up observations) and continues to resist classification, then the measurements are described and shared with the scientific community in scientific papers. Distinct groups of researchers using independently-developed instrument systems are invited to verify repeated observations of the candidate (should it reappear in the same part of the feature space at the same or other geographic locations). This task need not involve all of the preceding steps, since the characteristics of the phenomenon are by this time well-constrained and tailored instrumentation can be designed to focus on the characteristics associated with the candidate in question. A candidate that has resisted systematic attempts at classification and explanation through the foregoing process, and that does not represent a novel human technology that comes eventually to light, resides outside the "phenomenological envelope" of known phenomena and can be regarded as a thing previously unknown to science: a *bona fide* statistical anomaly.

(9) *Formulate and test hypotheses:* An investigation of the kind just described may uncover interesting statistical anomalies that nevertheless have prosaic explanations. An example is the elusive atmospheric phenomenon known as "sprite" formation, involving large-scale electrical discharges far above storm clouds (Liu *et al.*, 2015). An aerial census could have discovered sprites as a statistical anomaly (their properties reside outside of the phenomenological envelope of ordinary lightning, for example), but plausible explanations have been developed in terms of current scientific theory. If a statistical anomaly instead resists explanation in terms of prevailing scientific beliefs (i.e., prevailing scientifically-informed theories or dogmas) after rigorous investigation that involves hypothesis formulation and testing, then it may at last be recognized as a scientific anomaly.

The primary scientific objective of the UAP investigation consists of the steps described in this section: i.e., conducting the aerial census and categorizing phenomena in the census as (i) naturally-occurring phenomena, (ii) human-made phenomena, and (iii) "other," which is further subdivided into these subcategories: (a) statistical outliers, (b) corroborated outliers, (c) statistical anomalies, and (d) scientific anomalies. In the majority of detections, we expect it will be possible to confidently identify known objects and phenomena. Under some circumstances, such as where corroborated observations indicate that an object exhibits anomalous kinematics, we expect it could be possible to conclusively rule out an association with known phenomena. The system will be tested to verify these capabilities using synthetic data injection events, as well as spoofing events carried out using drones.

It should be acknowledged that this work cannot be used to prove the null hypothesis. That is, if we find no statistical or scientific anomalies, this does not imply that none exist. At best, we can place bounds on the frequency with which anomalies occur in the detection volumes that we examine and at the locations where we have installed instruments. A second application of this work is the evaluation of UAP reports about events that occur within the detection volume of our instruments. While this is not our primary concern, learning to evaluate reports may help us to leverage these informal data sets to inform decisions about where to locate our instruments. The census data can be used to evaluate contemporaneous UAP reports and shed light on the characteristics of objects that members of the public have identified as unusual, in some cases providing explanations for individual events.

## 4. Methodology for Defining Scientific Requirements

The Galileo Project will perform a wide survey of aerial phenomena, which includes measurements that may have few or no significant representatives in the UAP literature. The instrumentation suite is designed to cast a wide net with a fine mesh using parallel and simultaneous measurements with a high degree of self-corroboration. The "wide net" refers to large coverage of the sky from many locations and simultaneous wide-band coverage of the EM and acoustic spectrums over long time scales. The "fine mesh" implies high spatial, frequency, and time resolution. The answers to instrument-specific questions about the sufficiency of coverage and resolution have been decided in light of four principal considerations: (1) the performance envelopes of known phenomena; (2) the available data about UAP from UAP reports and studies of those reports; (3) physical limitations imposed on coverage, accuracy, and precision; and (4) practical limitations on the same.

(1) **Performance envelopes of known phenomena:** Measurement requirements can be informed by the necessity of measuring characteristics that surpass the capabilities of known human technologies or the kinematics of known natural objects. For example, if the swiftest acceleration of modern rockets is of order several hundred Earth gravities (e.g., $\sim 400g$ for the HIBEX missile (Atta *et al.*, 1991)), then a system designed to characterize kinematics must be at least capable of measuring accelerations significantly in excess of this value. This can help to establish the sampling rates required to measure the time rate of change of Doppler shifts using a radar system.

Some requirements identified in this way may have necessary correlates in order to be convincing; these are called *compound requirements*. For example, a low-flying object whose constant velocity is measured by radar to exceed the speed of sound is not by itself convincing of capabilities exceeding the performance envelope of known aircraft. On the other hand, if this object exhibits no thrust plume in thermal IR imagery and if no sonic boom is recorded by microphones, then its behavior is more convincingly unusual. This example illustrates why the ability to measure performance greatly in excess of average performance, even if not exceeding record performance, is nevertheless useful. Status as an outlier can be established by multiple features in combination; even moderately exceptional behavior in more than one dimension can distinguish an object as highly uncommon.

(2) **Characteristics of UAP from historical reports:** The set of available data about UAP can be leveraged to inform requirements in at least three ways: (i) by reference to the performance envelope of UAP; (ii) by reference to an archetypal or average UAP at a specific distance from sensors; (iii) by reference to scientific analyses of UAP reports and the measurements required to decide among hypotheses (i.e., to ensure the specific compound measurements missing from past reports are not missing from future measurements).

(i) *UAP performance envelopes*: This approach considers the extrema of reported UAP characteristics and identifies measurement requirements based on the ability to cover most or all of this envelope. Unfortunately, the aggregate of UAP reports do not supply upper bounds on many of the most extraordinary characteristics of UAP, such as their purportedly extreme speeds and accelerations or magnetic field strengths (Knuth *et al.*, 2019; Weinstein, 2012). In some cases, this leaves us with no clear answer to the question of what measurement range is appropriate.

(ii) *Archetypal or average UAP*: This approach relies on the properties of an archetypal UAP drawn from a class of well-documented sightings. The goal is to ensure that instrumentation, if located sufficiently close to the event, would make the necessary identifying detection or measurements.

Choosing a single archetype is difficult because the UAP described in reports have many manifestations and appear to belong to multiple categories of phenomena (e.g., nocturnal lights and daylight discs (Hynek, 1972a,b)). Moreover, individual UAP have been reported to dramatically change form or luminosity during a single sighting (Teodorani, 2004). The disadvantage of this approach, therefore, is that the assumed characteristics — the specific reference values of emitting power, size, speed, and acceleration — are chosen somewhat arbitrarily and are unlikely to be representative of the broad swath of reported UAP sighting scenarios. We can bracket this problem to some extent by using "average" or most frequently occurring characteristics as described in scientific and governmental compendia, to ensure that at least the "typical" characteristics are covered by our instrument specifications.

(iii) *Scientific analyses of UAP reports*: A third approach makes use of well-known scientific analyses of historical UAP reports that have drawn no firm conclusions on account of insufficient data. The goal in this case is to consider what observation(s) would have been required to confirm or rule out the null hypothesis, and to design the instrument system in a way that guarantees a successful determination were the circumstances of the scenario to be repeated.

(3) **Physical limitations:** In a few cases, there is a physical limit on the useful range or precision of measurements. For example, atmospheric turbulence limits the resolution of ground-based telescope observations to about 1 arcsecond/pixel. Atmospheric absorption of infrared light by water vapor and $CO_2$ restricts the useful band for detection of objects to between approximately 8 and $14\,\mu m$, and in the near to mid-wavelength infrared to between 0.2 and $5.5\,\mu m$. Absorption coefficients of sound in air increase with increasing frequency, resulting in maximum propagation lengths of less than 20 m for frequencies above 200 kHz.

(4) **Practical limitations:** When performing exploratory science of this kind, the answer to "how much coverage" and "how much resolution" is sufficient for measurements of individual observables is commonly "the greatest possible within the limits of project resources and current technology." This kind of justification is not uncommon in new fields of study, where relatively unconstrained searches are undertaken for phenomena whose properties are poorly understood. A prominent example is the ongoing search for dark matter, in which the properties of weakly-interacting massive particles (WIMPs) are poorly constrained by theory and the search is increasingly driven by advances in instrumentation technology and limited by financial resources (Mayet *et al.*, 2016; Billard *et al.*, 2022).

The Galileo Project instrument suite is designed with the consideration of expense in mind, so that multiple copies can be assembled and distributed to increase spatial coverage. For example, rather than purchase an expensive network of active radar stations, a custom passive radar system has been designed to track object kinematics, which can be replicated at a fraction of the cost of comparable active commercial systems. For another example, when deciding the rate at which mirrors should be capable of tracking an airborne object for narrow-field image capture, the answer from consideration of the UAP reports is unbounded; there is no confident upper bound of UAP velocity, and even relatively slow-moving objects at close range will require high angular velocities for tracking. The fastest possible sweep rate is determined by the servo and control system, and cost and required design effort increases dramatically for angular speeds exceeding roughly 180 deg./s. This is an example of a resource constraint that dictates the limit of a measurement capability.

Finally, we should note that limitations imposed by resource constraints lead in some cases to an unavoidable prioritization of measurements or the identification of acceptable trade-offs. For example, the short wavelength infrared (SWIR, 900–1700 nm) is currently not covered by our camera suite, and could be used to increase detection range under hazy conditions (Hansen & Malchow, 2008). Unfortunately, because wide-field SWIR cameras are very expensive (upwards of \$100 k), we have focused instead upon the long wavelength infrared (LWIR, $8–14\,\mu m$) where thermal emitters in the terrestrial atmosphere are bright enough for detection, and for which inexpensive uncooled cameras are readily available.

## 4.1. *Summary of procedure for deciding measurement requirements*

Because one of our primary objectives is to characterize phenomena that are very poorly understood, we shall perform a broad survey targeting the largest range in energy of acoustic, electromagnetic, and particle emissions, by employing all of the principal measurement modalities, including spectroscopy, imaging, and radar. For each combination of observable (e.g., time rate of change of Doppler shift) and physical parameter (e.g., acceleration), we consider the four factors that inform our measurement requirements as described above. First, if there is a useful quantity from the performance envelope of known phenomena, this dictates usually a lower bound for the range of required measurements (e.g., $400g$ for the case of acceleration).

Second, we consider scientific analyses of historical UAP reports to determine what range and precision of measurements would have been required to make a useful determination about the nature and origin of historical UAP observations. If, as typically happens, the measurement requirement after these considerations remains unconstrained (e.g., at least $400g$ but as high as possible), then it is determined through consideration of physical and practical limitations.

## 5. Science Traceability Matrix

The science traceability matrix (STM) is the foundational, organizing document of a complex scientific investigation. In preparing the STM for the Galileo Project's UAP study, we have approximately followed the guidelines of the NASA PI Launchpad seminars on science traceability and proposal writing (Feldman, 2019). An important difference concerns the manner in which requirements have been decided (see Sec. 4) because the characteristics of the phenomena we seek to study are not well defined. We should also note that the STM is a living document, and is likely to evolve through future stages of project development. Table 2 provides the project goals (which are aspirational), as well as primary objectives (which are likely achievable). Project-level requirements are also summarized, where these apply to the study as a whole, as opposed to individual instruments. These are followed in Tables 2–5 by instrumentation requirements, in the form of physical parameters and observables (see Sec. 3 for definitions) and their related measurement requirements. In the remainder of this section, we discuss the project-level and instrument measurement requirements in greater detail. A description of individual instruments and projected performance is saved for Sec. 6.

### 5.1. *Project-level requirements*

The requirements listed below are essential for meeting the project goals and objectives, as well as for the calibration of instruments and the acquisition and interpretation of instrument data.

(1) *Development, testing, and assembly of integrated instrument systems*: The foundational project-level requirement calls for the development of integrated instrumentation for sustained monitoring of the sky and environment using multiple measurement modalities.

(2) *Laboratory and on-site calibration of instruments*: Practically all instruments require at least some laboratory calibration (e.g., intrinsic calibration of all-sky camera optics to measure lens distortion (Szenher, 2023)) and some on-site calibration (e.g., extrinsic calibration of cameras to measure their orientation in a world coordinate frame). Additional examples include magnetometer readings of the geomagnetic field; and the intensity levels measured by the thermal IR cameras, calibrated using targets of known temperature.

(3) *Site selection and characterization*: Multiple locations will be selected for siting instruments for continuous operation over at least five years. Sites will be selected according to their suitability for instrument development, testing, and final deployment (continuous autonomous operation). This work involves a detailed characterization of the physical setting in which measurements are acquired (e.g., weather, typical atmospheric conditions, topography, and effective horizon). Important considerations relevant to site selection are discussed in Sec. 7.

(4) *Weatherization of instruments and automation of data collection*: All instruments at the time of deployment must be protected to endure harsh weather conditions including rain, snow, extreme temperatures, and high winds, and must operate unattended for 24 h each day and throughout the year.

(5) *Appropriate computer hardware and data storage infrastructure*: To accommodate the enormous data volumes generated by our instruments, we require computing hardware to encode and record data directly from our instruments, and in some cases to perform a first-pass data reduction. Meeting the project objectives will also require local GPU-enabled data servers for second-pass data reduction and analysis, as well as copious cloud storage to save event data (see Cloete (2023) in this volume for estimates). Following the completion of a comprehensive aerial census, only event data with a high aggregate outlier score will be archived. A detailed discussion of computer hardware and data storage requirements can be found in (Cloete, 2023, this volume).

(6) *Monitoring of environmental conditions*: The local environmental conditions will be continuously recorded using on-site instrumentation (e.g., temperature, pressure, humidity, dew point, cloud cover, wind velocity). Additional regional environmental data required for understanding large-scale atmospheric conditions (such as wind velocity at high altitude) can be accessed from online archives curated by NOAA (National Oceanic and Atmospheric Administration).

(7) *Monitoring of public air and space traffic reports*: The positions and velocities of nearby objects reported by independent sources will be continuously recorded. This includes logging of air traffic using ADS-B transponder messages (Automatic Dependent Surveillance-Broadcast) with an on-site receiver, as well as commercial and scientific satellite overflights and rocket launches using internet services devoted to this purpose. Continuous ADS-B recording will be required also for calibration of wide-field cameras and for training inference models to recognize conventional aircraft in imagery and acoustic data records.

(8) *Consultation of regional sensor networks*: Some instrument data records should be compared with readings from regional sensors that are continuously posted online: e.g., regional networks of infrasound monitors and magnetometers. This is essential for distinguishing events with effects that are purely local from those whose effects are regional.

(9) *Recording the astronomical context of observations*: Recording the astronomical background is essential for camera calibration and for rejecting detections related to known astronomical objects. These data include: (i) the ephemerides of the Sun, Moon, and visible planets, including rise and set times; (ii) windows of high meteoritic activity; (iii) the expected duration of astronomical twilight.

## 5.2. *Instrument requirements*

The STM is shown in Tables 2–5. Unless marked as "Phase 2+," performance requirements and projected performance refers to an instrument or capability that is included in Phase 1: i.e., the first development phase of our study. The bold-faced word "targeted" refers to observations that target a specific region of the sky where an object has been detected in wide-field imagery or radar. Targeted observations require tracking of objects as they move across the sky using a pan-tilt camera or telescope mount (Phase 1) or a pan-tilt mirror-tracking system (planned for Phase 2). "Phase 2+" refers to subsequent development phases, which will continue while Phase 1 instrumentation is tested and deployed. For completeness, we have mentioned approaches that may be employed in Phase 2 and beyond, but we leave this for future work and do not describe these requirements in detail. Items relevant to later development phases have been lightly shaded in Tables 2–5.

The majority of the physical parameters and observables listed in Tables 2–5 are time-dependent and our goal in most cases is to measure this variation with the highest sampling rates that are practical given cost and data storage constraints, as well as instrument limitations such as latency, frame rate, and calibration requirements. This optimization is necessary in light of the transient nature of events described in UAP reports, with properties such as brightness and position that in some cases vary on short time scales (∼1 s or less, Teodorani (2004)).

### 5.2.1. *Appearance and surface features*

The shape, size, and color of aerial objects and of their surface features are of vital importance for identification. Many aerial objects will exhibit features that indicate a means of propulsion (jets, rotors, propellers) or determine flight characteristics (wings or fins); the absence of such features is commonly noted in UAP reports (Condon, 1969; Sagan & Page, 1972; UKMoD, 2006). We therefore require cameras with high angular resolution that are able to capture objects crossing the sky with potentially very high angular velocity (tens of degrees per second). The mount and tracking mechanism should mitigate jitter, which can lead to motion blur and loss of detail. High frame rates are preferred for the same reason.

Most objects will be detected at low elevation angles, and are captured through a significant air mass, which degrades the effective maximum angular resolution to ∼1 arcsecond owing to refractive distortion from atmospheric turbulence. Narrow-field tracking cameras should therefore be chosen to maximize angular resolution up to 1 pixel/arcsecond.

Finally, measuring the Stokes parameters $(S_0, S_1, S_2, S_3)$ using polarimetry and polarimetric imaging is useful for discriminating smooth from rough surfaces (e.g., metallic from rough matted-like

Table 2. Science Traceability Matrix part 1. Shaded boxes refer to instrumentation projected for later project phases (i.e., Phase 2+).

**Goal:** To determine whether there are measurable phenomena in or near Earth's atmosphere that can be confidently classified as **scientific anomalies.**

**Objectives:** 1. *Detect, characterize, and catalog* aerial phenomena with sufficient accuracy to distinguish between: (i) known human-made objects (e.g., known human technology and other human artifacts); (ii) known naturally-occurring phenomena (e.g., atmospheric, biological, geophysical, astrophysical, etc.); (iii) phenomena that fit neither of the preceding categories: outliers, statistical anomalies, and scientific anomalies. 2. *Archive measurement data* for evaluation by the Galileo Project and, following internal team validation of the data, the broader scientific community.

**Project-level requirements:** 1. Development, testing, and assembly of integrated instrument systems; 2. Laboratory and on-site calibration of instruments; 3. Site selection and characterization; 4. Appropriate computer hardware and data storage infrastructure; 5. Weatherization of instruments and automation of data collection; 6. Monitoring of environmental conditions; 7. Monitoring of air and space traffic reports; 8. Consultation of regional sensor networks; 9. Recording the astronomical context of observations.

| | Physical parameters | Observables | Instrument Type and Performance Requirements | Instrument Model and Projected Performance |
|---|---|---|---|---|
| **Appearance & surface features** | **Morphometry:** overall *shape* and *size*; *shape* and *size* of surface features (e.g., wings, jets, rotors, structural elements) | Optical narrow-field color imagery | **Targeted:** Narrow-field optical camera; maximize angular resolution up to 1 pixel/arcsecond: e.g., resolves features measuring 5 cm in diameter at 10 km range; jitter $\ll$ 1 arcsecond | Beacon 8.0 surveillance camera: 8 MP at 30 fps; 2-60 arcsecond/pixel with optical zoom |
| | | | | TBD (Phase 2++): 1 arcsecond / pixel |
| | | Infrared narrow-field imagery | **Targeted,** Phase 2+: narrow-field infrared cameras; *maximize angular resolution* | TBD (Phase 2+) |
| | **Surface roughness** [m] | Polarimetrically-derived Stokes parameters as a function of wavelength | **Targeted,** Phase 2+: narrow-field polarimeter or polarimetric imager | TBD (Phase 2+) |
| **Position** | **Angular position**: *azimuth*: 0º to 360º; *altitude*: 0º to 90º. | Optical, UV, and infrared wide-field imagery | All-sky optical camera; *maximize resolution* and *frame rate* | Alcor OMEA 9C all-sky camera: 32 MP at 3 fps; 180º × 180º FOV |
| | | | | ZWO ASI462MC optical+NIR camera: 2.1 MP at 136 fps; 170º × 170º FOV; 400-1000 nm |
| | | | | NPACKMAN all-sky camera: Sony Exmor RS CMOS IMX477 sensor (12.3 MP, 400-700 nm); 170º × 170º FOV; 15 fps |
| | | | All-sky UV-sensitive camera; *maximize angular resolution* and *frame rate* | TBD (Phase 2+) |
| | | | All-sky array of LWIR wide-field cameras; *maximize angular res. and frame rate* | 8 FLIR Boson uncooled microbolometers (640x512 px, 8 $\mu$m-14 $\mu$m); 7×50º+1x95º FOV |
| | | Optical and infrared narrow-field imagery for position refinement | **Targeted:** Narrow-field optical camera with high angular resolution | Beacon 8.0 surveillance camera: 8 MP at 30 fps; 2-60″/px with optical zoom |
| | | | **Targeted,** Phase 2+: narrow-field optical & infrared cameras with high precision pointing; *maximize angular resolution* (1″/px) | TBD (Phase 2+) |

Table 3.   Science traceability matrix part 2.

| | | | | |
|---|---|---|---|---|
| **Position** | | Acoustic direction finding from time delay of cross-correlated audible signals | Phase 2+: local array of 4 audible-band microphones | TBD (Phase 2+) |
| | | Radio direction finding from time delay of cross-correlated radar echoes (from single location) | Phase 2+: local array of 4-5 antennas with phase-coherent receivers | TBD (Phase 2+) |
| | **Geographic position**: *latitude* [deg.], *longitude* [deg.], *elevation* [km]<br><br>**Range**: distance to the object [km] | Time delay of cross-correlated radar echoes at multiple receivers to triangulate position | Regional multistatic passive radar: regional array of radio receivers; *maximize frequency resolution* and *minimize noise floor* to improve range and precision of position estimates | SkyWatch multistatic passive radar system (GP-developed); frequency resolution: 1 Hz-16 Hz (far field-near field); detection range: airliner (150 km, 20 dBsm); light aircraft (110 km, 15 dBsm); bird (6 km, -35 dBsm) |
| | | Optical and IR triangulated positions from multiple simultaneous all-sky images acquired from endpoints of a ~several km baseline | Regional array of optical and IR cameras; *maximize resolution, frame rate, temporal accuracy & precision*, and *minimize latency* (for synchronization) | Additional all-sky cameras (see Alcor OMEA 9C and ZWO ASI462MC, NPACKMAN all-sky camera, and FLIR Boson 640 specifications listed above) |
| | | Acoustically-triangulated positions from cross-correlated audible signals from multiple microphones | Phase 2+: regional array of audible band microphones | TBD (Phase 2+) |
| **Kinematics** | **Translational:** *velocity* [m/s]; *acceleration* [m/s²] | Radar echo Doppler shift, rate of change of radar-derived position (see *position*); rate of change of Doppler velocity | Regional multistatic passive radar: regional array of radio receivers; *maximize frequency resolution* and *minimize noise floor* to improve range and precision of velocity and acceleration estimates | SkyWatch multistatic passive radar system (GP-developed); frequency resolution: 1 Hz-16 Hz (far field-near field); max accelerations: $150g$-$2800g$ (err. $\pm 0.15g$-$4.5g$); max Doppler velocity: 1.5-1.7 km/s (error $\pm 1.5$-2.8 m/s) |
| | | Optical Doppler shift (for velocities >10 km/s) | **Targeted**, Phase 2+: high-res. VIS/NIR spectrometer | TBD (Phase 2+); $\lambda/\delta\lambda \geq 30{,}000$ for velocity $\geq 10$ km/s |
| | | Acoustic Doppler shift measured using multiple microphones | Phase 2+: regional array of audible band microphones | TBD (Phase 2+) |
| | | Sonic boom (distinctive waveform and power spectrum); implies supersonic speed | Infrasonic and audible band microphone | GRAS Acoustics 41AC-3 CCP audible mic (10 Hz-20 kHz); Infiltec INFRA20 infrasound monitor (0.05 Hz-20 Hz) |
| | **Rotational:** *rotation period* [s] and *rotation axis orientation* [deg. × deg.] | High-speed narrow-field imagery | **Targeted:** High-speed narrow field camera; *maximize frames per second* | Beacon 8.0 surveillance camera: 8 MP at 30 fps; 2-60 arcsecond/pixel w/optical zoom<br>TBD (Phase 2+): e.g., i-Speed 7 Olympus camera, 3.2 MP, 5k fps |
| | | Polarimetry (period of polarization of reflected sunlight) and photometry (period of brightness modulation) | **Targeted**, Phase 2+: high-speed photometer and polarimeter (or polarimetric imager); *maximize fps* | TBD (Phase 2+) |
| | | Doppler broadening of spectral lines | **Targeted**, Phase 2+: optical and IR spectrometers; *maximize frequency resolution and SNR* | TBD (Phase 2+) |
| | | Doppler frequency dispersion of radar echoes | Phase 2+: radio receivers; *maximize frequency resolution* | TBD (Phase 2+) |

surfaces); most natural surfaces are optically rough. The reflected polarization from an object or surface varies as a function of surface roughness, and other factors such as albedo and the geometry of the observation (Schott, 2009). Thus, polarimetry or polarimetric imaging is very useful for detecting artificial or man-made objects against natural backgrounds, since these objects tend to produce specular responses and thus an enhancement of the return polarization response under an appropriate viewing geometry.

Table 4. Science traceability matrix part 3.

| Material properties | **Composition** (elemental, molecular) | Optical, IR, UV, radio & microwave spectra (emission and reflectance) and derived spectral parameters: e.g., absorption bands and lines, emission bands and lines, continuum slope, band ratios | **Targeted**, Phase 2+: optical, infrared, and ultraviolet spectrometers; *maximize frequency resolution, SNR* | TBD (Phase 2+): spectrometers and imaging spectrometers; λ/δλ ≥ 500 |
|---|---|---|---|---|
| | **State** (solid, liquid, gas, or plasma) | | **Targeted**, Phase 2+: wide-band directional antennae & receivers; *maximize frequency resolution* and *SNR* | TBD (Phase 2+): |
| State variables | **Pressure** [Pa] | Plasma: pressure broadening of optical emission lines | **Targeted**, Phase 2+: optical spectrometer | TBD (Phase 2+) |
| | | Gas: pressure broadening of absorption bands in IR transmission spectra | **Targeted**, Phase 2+: IR spectrometer | TBD (Phase 2+) |
| | **Density** [kg/m³] | Solid: optical reflectance spectrum of surface material (of known density) identified using spectral library; i.e., used to infer density of surface material(s) | **Targeted**, Phase 2+: optical spectrometer | TBD (Phase 2+) |
| | **Temperature** [K] | Optical and infrared apparent brightness as a function of wavelength; see *luminosity* | Intensity-calibrated wide-field LWIR cameras; *maximize dynamic range, accuracy* and *precision of intensity measurement* | 8 FLIR Boson microbolometers (640x512 px, 8 μm-14 μm); 7×50º+1×95º FOV |
| | | | **Targeted**, Phase 2+: IR spectrometer: ratio of incident flux from 1-5 μm and 7-14 μm | TBD (Phase 2+) |
| | | Temperature-dependent optical electronic emission and absorption spectra | **Targeted**, Phase 2+: optical spectrometer | TBD (Phase 2+): λ/δλ ≥ 500 |
| Optical and infrared brightness (and radiometric coefficients) | **Luminosity** (wavelength dependent)**:** overall *optical* and *IR luminosity* [W]; and *characteristic optical and infrared emission spectra* [W/nm]<br><br>See also: implications for *composition*, *state*, and *temperature*. | Intrinsic irradiance in nighttime optical and daytime + nighttime IR, band-integrated (photometric) and as a function of wavelength (spectrometric); also requires emitter *size, shape*, and *range* | Intensity-calibrated wide- and narrow-field optical and IR cameras; *maximize dynamic range, frame rate, accuracy* and *precision of intensity measurements* | See Alcor OMEA 9C and ZWO ASI462MC, and NPACKMAN all-sky camera specifications listed above.<br>8 FLIR Boson microbolometers (640x512 px, 8 μm-14 μm); 7×50º+1×95º FOV<br>TBD (Phase 2+): High fps camera for high-speed photometry: e.g., i-Speed 7 Olympus, 3.2 MP, 5k fps |
| | | | **Targeted**, Phase 2+: IR spectrometer: ratio of incident flux from 1-5 μm and 7-14 μm | TBD (Phase 2+) |
| | **Reflectivity** and **emissivity** | Apparent brightness in daytime and nighttime optical + IR as a function of wavelength | **Targeted**, Phase 2+: optical and IR spectrometers | TBD (Phase 2+) |

### 5.2.2. *Position and kinematics*

Knowing the distance to an object is important for estimating many useful quantities such as the object dimensions, velocity, and luminosity. In addition, one of the most common themes in UAP reports is the description of objects that appear to undergo enormous accelerations (Knuth *et al.*, 2019).

Conventional aircraft have a performance envelope that depends on atmospheric density and hence on altitude (see Fig. 1); objects performing outside of this envelope of course merit close scrutiny. Objects at high altitude (>30 km) that remain aloft while moving at speeds intermediate between strato-spheric winds and orbital velocities should be regarded as highly unusual. For all of these reasons, estimating object position, velocity, and acceleration is of paramount importance. Of special interest is the ability to measure velocities well in excess of conventional aircraft at several multiples of the sound speed in air ($\sim 0.34$ km/s at sea level, 15°C), and accelerations exceeding the highest-accelerating rockets at hundreds of Earth gravities (e.g., HIBEX,

> 400*g*; Atta *et al.* (1991)). Range estimates that can reach up to several tens of kilometers are necessary to reach the upper atmosphere.

Table 5. Science traceability matrix part 4.

| | | | | |
|---|---|---|---|---|
| **Sound** | **Sound source pressure level** (frequency dependent): *Characteristic acoustic emission spectrum* [Pa/Hz] with implications for: *propulsion mechanism, structural resonance, turbulence, wakes, shock wave emission*; see also: implications for *velocity* | Power spectrum in multiple bands: infrasonic (0.05 Hz - 20 Hz), audible (10 Hz - 20 kHz), and ultrasonic (16 kHz - 190 kHz); also requires *range* | Infrasonic, audible, and ultrasonic microphones; *maximize sensitivity, dynamic range, and SNR* | Infiltec INFRA20 infrasound monitor (0.05 Hz-20 Hz) |
| | | | | GRAS Acoustics 41AC-3 CCP audible mic. (10 Hz-20 kHz) |
| | | | | Wildlife Acoustics SM4BAT with SMM-U2 microphone (16 kHz to 190 kHz) |
| | | | **Targeted**, Phase 2+: directional wide-band audible band microphones | TBD (Phase 2+) |
| **Radio** | **Communication-related radio transmission** [dBm/Hz] **:** VHF (30 MHz - 300 MHz), UHF (300 MHz - 3 GHz) | Modulated signals in communication bands in VHF and UHF reserved for air traffic and remote controlled vehicles | Radio receiver and spectrum analyzer with wide-band antenna; *maximize bandwidth, frequency resolution, SNR, and sweep rate* | Signal Hound BB60C Spectrum Analyzer (9 kHz-6 GHz); 10 Hz-10 MHz res. bandwidths (RBWs); 24 GHz/s sweep rate @ ≥10 kHz RBW; discone antenna (75 MHz-3 GHz) |
| | | | **Targeted**, Phase 2+: directional wide-band antenna and receiver | TBD (Phase 2+) |
| | **Other characteristic radio and microwave emission** [dBm/Hz]; see also implications for *state* | Wide band power spectrum of radio and microwave signals | Radio receiver and spectrum analyzer with wide-band antenna; *maximize bandwidth, SNR, sweep rate.* | SignalHound BB60C Spectrum Analyzer (see specifications above) |
| | | | **Targeted**, Phase 2+: directional wide-band antenna and receiver | TBD (Phase 2+) |
| **Energetic particles** | **Characteristic particle emission**: *counts per second* [cps], *particle energy* [MeV] | Particle counts per second with energies exceeding 1 MeV | Energy-calibrated scintillation counter with photomultiplier | NPACKMAN integrated scintillation counter with coincidence mode for muons. Sensitive to gamma radiation (100 keV to 5 MeV), fast neutrons and charged particles. |
| | | Particle energy derived from magnitude of photomultiplier voltage signal (adjacent to scintillator) | | |
| | | | **Targeted**, Phase 2+: directional detector | TBD (Phase 2+) |
| **Quasistatic fields** | **Magnetic dipole moment** [Am²] and **orientation** [deg. × deg.]; or **equivalent current density** [A/m²] | Local magnetic field intensity and direction; also requires *range* | Magnetometer; resolution of order ~nT to resolve diurnal geomagnetic field variations; measurement range should encompass overall geomagnetic field variation. | Bartington Mag-13MS100; resolution: ~1 nT; accuracy: few nT; measurement range: ±100 $\mu$T |
| | | | | Modified triaxial FG3+ fluxgate magnetometer; resolution: ~1 nT; measurement range: ±50 $\mu$T |
| | | Zeeman effect splitting of emission lines in optical spectra to measure field strength at source | **Targeted**, Phase 2+: optical spectrometer; *maximize frequency resolution, SNR* | TBD (Phase 2+); $\lambda/\delta\lambda \geq 2500$ for 10 T field |
| | **Electric charge** [C] | Local electric field (V/m); also requires *range* | Phase 2+: electrostatic field meter | TBD (Phase 2+) |
| | | Stark effect splitting of emission lines in optical spectra to measure electric field strength at source | **Targeted**, Phase 2+: optical spectrometer; *maximize frequency resolution, SNR* | TBD (Phase 2+) |

Positions are estimated using triangulation of detections measured via optical and IR image capture, radar echoes, and acoustic recordings. In Phase 1, we address only optical and IR image triangulation techniques and triangulation of radar echoes (Szenher, 2023; Randall, 2023, this volume). All approaches require very precise timekeeping and correlation at multiple instrument sites. Velocity can be measured directly using acoustic, optical, and radar Doppler shifts recorded at multiple ground stations; we will focus on radar Doppler in Phase 1. High resolution optical spectrometers ($\lambda/\delta\lambda > 30,000$) can be used to measure optical Doppler shifts for velocities in excess of 10 km/s, and may be incorporated in Phase 2 or beyond. The Doppler shift of an aircraft-related sound, which can be measured by a single microphone and used to estimate radial velocity, is illustrated in Mead (2023) in this volume. Supersonic velocities are indicated where individual microphones measure sonic booms.

Finally, rotational motion can be estimated from narrow-field imagery acquired at high frame rates. In later development phases we shall consider approaches that make use of high-speed polarimetry and photometery, as well as the Doppler velocity dispersion of radar echoes. Although challenging, it may also be feasible to use high-resolution optical and infrared spectrometers to measure Doppler line broadening and in this way detect extreme rotation rates.

### 5.2.3. *Brightness*

The wavelength-dependent luminosity of aerial objects depends on many factors that are challenging to independently constrain. In Phase 1 our requirement is to make a simple estimate of luminosity using the intrinsic irradiance measured with wide-field cameras in the long-wavelength infrared band (LWIR, $8\,\mu m$ to $14\,\mu m$), in which atmospheric infrared transmissivity is relatively high. This estimate assumes the object in question is a blackbody emitter, and must take account of atmospheric humidity and elevation angle (and associated air mass) which can affect the measured flux. This luminosity estimate also requires an estimate of the range to the emitter as well as its size and approximate shape. In a later development phase, a more accurate estimate can be achieved using a wide-band infrared spectrometer to ratio the incident flux in the $1$–$5\,\mu m$ and $8$–$14\,\mu m$ bands, and in this way estimate the peak emission wavelength. The use of optical and infrared spectrometers to estimate radiometric coefficients such as reflectivity and emissivity is postponed until a later development phase.

### 5.2.4. *State variables and material properties*

Determining material properties such as composition and physical state, and state variables such as pressure, temperature, and density, largely depends on spectroscopic observations. The use of spectrometers (including imaging spectrometers) has been postponed until Phase 2 of our study, but we have included a mention of these approaches in the STM for completeness (Tables 2–5). In particular, the presence and wavelengths of specific emission and absorption features in optical, UV, and infrared spectra can be used to distinguish between states of matter and specific compositions in gases and plasmas. Also for gases, broadening of spectral features can be used to infer pressure and temperature and, when combined with equations of state, can be used to infer density. The only state variable that we will measure in Phase 1 is the blackbody surface temperature, which is inferred from the estimated luminosity (see Sec. 5.2.3).

### 5.2.5. *Sound*

In the majority of UAP reports, sightings have been associated with a notable *absence* of sound. Where sound is reported, it is not typically associated with conventional propulsion systems (Sagan & Page, 1972; Mead, 2023). Sound is commonly described as having low frequency and, in the case of some proximate sightings, high-frequency sounds are noted instead or in addition (Johnson & Saunders, 2002; UKMoD, 2006).

Sounds produced by conventional airborne vehicles have intrinsic as well as extrinsic causes. Intrinsic sources relate to sounds generated by internal mechanisms, such as the enclosed components of aircraft engines that produce sound even while the vehicle is stationary. Extrinsic sources arise from the interaction of the vehicle frame and propulsion system with the surrounding atmosphere. Buffeting of the frame excites vibrations that also generate sound. Sonic boom is an extrinsic shock production mechanism that arises from translation through the atmosphere at supersonic speeds. Lightning also produces shock waves from rapid expansion of super-heated air. Finally, airborne animals produce sounds that are readily recognizable, including complex audible calls and songs (in birds) and echo-ranging chirps (in bats), as well as the low-frequency thrum of wing flaps.

The power spectra of sounds generated by all of these phenomena are distinctive and can be identified using well-established analytic methods, provided they are measured with sufficient accuracy and precision, and the signal exceeds the ambient noise level. The baseline requirement for acoustic measurements, therefore, is to use calibrated, high-precision microphones (measurement microphones) to record sound pressure level as a function of frequency and time. When combined with range, an understanding of the ambient noise field, and transmission path to the sound source, acoustic recordings allow us to infer a sound emission spectrum for frequencies that have not been attenuated to extinction by passage through the atmosphere.

We have selected our acoustic measurement instruments to cover the infrasound through ultrasound frequency bands. (Ultrasonic waves attenuate rapidly in air so that effective detection volumes are small in this band.) Instruments with high sensitivity, high dynamic range, low intrinsic electronic noise floor, and a flat frequency response are ideal. This requires a high quality of manufacture that involves adequate shielding of the electronics from ambient sources of interference and noise. We shall focus on omnidirectional recordings in Phase 1; targeted or directional sampling will be considered for Phase 2. A detailed description of the Phase 1 acoustic instruments can be found in Mead (2023) in this volume.

### 5.2.6. *Energetic particles*

The motivation for including particle counters derives primarily from the need to characterize the total radiation environment and local effects of space weather upon instrumentation. This is an essential part of understanding the global conditions in which artifacts or unusual phenomena may arise. An additional but secondary motivation derives from rare reports of residual environmental radiation following UAP events and reported burns and other injuries related to high-energy EM or particle radiation (UKMoD, 2006). Until Phase 2, our efforts will focus on measuring particle counts per second and estimate particle energies, after which the ability to measure direction may be added. Our primary interest is to identify any marked changes in the particle flux that depart from the terrestrial and cosmic background of high-energy subatomic particles (muons, as well as alpha particles, electrons, protons, and neutrons) and gamma rays.

### 5.2.7. *Quasistatic fields*

Ferromagnetic objects and electrical currents generate magnetic fields. The intrinsic physical parameters in this case are the magnetic dipole moment and orientation, or the equivalent current density. Magnetometers can be used to measure the local geomagnetic field and detect geomagnetic storm activity, which provides additional valuable context for the interpretation of instrument readings. Secondarily, several studies from the historical literature investigated reports of magnetic disturbances in the presence of UAP, including: (i) sightings by military and civilian pilots who have reported strong perturbations of the onboard magnetic compass during an encounter (Weinstein, 2012); (ii) observed spikes in magnetometer readings that are coincident with UAP sightings and purported magneto-optical effects (Meessen, 2012); (iii) ground magnetic gradiometry anomalies that reportedly persist for a few days following a UAP sighting (Maccabee, 1994); and (iv) magnetic perturbations in a region where UAP sightings are relatively frequent (Akers, 2007).

The choice of the GP magnetometers is guided by the goal of reliably recording magnetic field perturbations that are associated with known and unknown sources at the instrument site. This is not trivial considering that the magnetometers will be collecting measurements in the field, throughout the year, at locations that will be magnetically noisy, and in the presence of geomagnetic field variations.

Examples of known physical sources include: the ionospheric solar quiet currents, which are responsible for the diurnal variation of the geomagnetic field in the order of tens of nT; the magnetospheric ring current during geomagnetically quiet times, which generates a signal of few nT up to few tens of nT; and geomagnetic storms, which can create perturbations of several hundreds of nT. Expected magnetic perturbations from human-made sources include signals from power lines (on the order of few nT to few tens of nT within 50 m distance, Garrido et al. (2003)), electric railways (on the order of 10 nT at few kilometers distance; Ding et al. (2021); Jankowski & Sucksdorff (1996)), airliners (of order a few nT at ~100 m distance) and motor vehicles like cars and trucks (of order a few nT within hundreds of meters) (Lenz & Edelstein, 2006; Chulliat et al., 2009).

Given the above, vector magnetometers have been selected that fulfill the following requirements: (i) the measurement range must exceed the ambient geomagnetic field range (i.e., exceeding $\pm 60\,\mu\mathrm{T}$); (ii) the noise level should be less than $1\,\mathrm{nT}\ \mathrm{rms}/\sqrt{\mathrm{Hz}}$ at 1 Hz; (iii) normal performance should be expected over the entire temperature range anticipated throughout the year at the instrument site; and (iv) the accuracy and resolution should be at least of order a few nT. The eventual acquisition of a high-resolution optical spectrometer in Phase 2+ may enable measurement of Zeeman effect splitting of specific emission lines to measure magnetic field strength at the source (Zeldovich et al., 1983). We estimate that a spectral resolving power of $\lambda/\delta\lambda \sim 2{,}500$ is sufficient to detect a very strong (10 T) field using, for example, the Hydrogen Balmer lines, Oxygen and Na D-lines (e.g., following the calculation in Mathys et al., 1997). Astronomers routinely use the Zeeman effect, as well as other magneto-optical effects that influence polarization and spectral lines, to characterize the magnetic fields of astrophysical sources (e.g., Hyder, 1965; Western & Watson, 1984; Stenflo, 2013). While it has been suggested that macroscopic magneto-optical phenomena ascribed to UAP may be detectable using optical cameras (Meessen, 2012), we will save exploration of this possibility for a later project phase.

Objects with nonzero charge produce electric fields. An electric field meter at a known range can be used to estimate the amount of charge at the source. We have included this in our plans for Phase 2 mainly for completeness, since verified high intensity electric field effects are not commonly noted in UAP reports. Also in Phase 2, the acquisition of a high resolution optical spectrometer may enable measurement of Stark effect splitting of emission lines to estimate electric field strength at the source (Harmin, 1982).

# 6. System Overview

In this section, we provide a high-level description of the instrument systems planned for our investigation. First we describe the classes of systems that we plan to develop over the long term. Our development process will consist of at least two phases. At time of submission, Phase 1 instruments have been assembled; testing of the instruments is underway, along with development of data processing pipelines. We provide an overview of this "observatory-class" system following a description of the other two classes ("portable" systems and "mesh" systems). Phase 2 will involve additional development over the coming years, including especially the mirror-tracking capability for narrow-field observations. Whereas the discussion in Sec. 5 was concerned with physical parameters and instrument requirements, the current section is concerned with describing the Phase 1 instruments and their expected performance (see the right-most column of the STM in Tables 2–5).

## 6.1. *System classes*

The Galileo Project is developing three classes of instrument systems for the investigation of UAP, each of which has a different role and addresses this challenge with different trade-offs. Table 6 summarizes the categories of planned systems, although the remainder of this paper will focus on an overview of the observatory-class system that has been assembled to date.

### 6.1.1. *Observatory systems*

Observatory systems are assembled from high-end scientific instruments optimized for accuracy, sensitivity, coverage, and resolution. Most instruments are factory calibrated. These systems will be deployed at sites with ample power and internet throughput, and can host servers with high-performance computing hardware for second-pass data reduction before transfer to cloud servers. These systems are designed for long-term deployment at a specific location and must function autonomously 24h/7d. They are also weatherized for continuous operation in most climate regimes. The all-sky camera arrays in observatory systems are designed to monitor the sky and environment in a radius of at least 15 km for objects 30 m in scale. (In practice, each instrument has its own detection volume which depends on the types and sizes of objects and the measurement modality.) Data reduction must happen on site and in real time to keep pace with the enormous data acquisition rates. Observatory class systems will be supplemented with nearby "secondary" stations (with ∼several km separation, Szenher (2023)) with redundant sensors as well as additional cameras and antennas for triangulating aerial object positions and kinematics, and for corroborating observations recorded at the primary stations. Observatory-class systems will require infrequent maintenance according to a schedule that is to be determined. All of the requirements that have been described so far in this paper and the STM refer to observatory systems.

Table 6. Instrument system classes.

|  | Observatory | Portable | Mesh |
|---|---|---|---|
| weatherization | total | partial | total |
| continuous operation | $\leq$ 1 yr | $\leq$ 2 wks | $\leq$ 1 yr |
| automation | total | partial | total |
| cost | ~\$250 k | ~\$25 k | ~\$2.5 k |

### 6.1.2. *Portable systems*

Portable systems will be optimized for easy portability and use at remote locations without access to utility power or hard wired internet connectivity. Portable systems are designed for rapid deployment and continuous operation for up to two weeks. These are monitored by an operator who attends to the instruments on a daily basis for the duration of a field campaign. Data is recorded and stored on portable media for later processing on GPU-equipped servers. Power is supplied by batteries that are charged using solar panels, although utility power can be used where available. To reduce cost and weight, portable systems are not necessarily weatherized, and are designed for operation in clement conditions.

### 6.1.3. *Mesh systems*

Mesh systems will be optimized for cost, so that many copies can be distributed over a large area (a network or "mesh"). These consist of low-end, consumer-grade sensors and instruments that will require lab calibration in the case of some sensors. Mesh systems will rely on a local solar battery power source or utility power for continuous operation over long time scales (up to a year) and are fully weatherized. Mesh systems are designed to operate as part of an extensive regional network whose goal is to triangulate object positions and and derive their kinematics. Each node will monitor the sky within a radius of up to 5 km (i.e., detection of a 30 m object at 5 km range). These systems will be designed to require minimal maintenance so that they can be hosted by non-experts. Mesh nodes will consist of omnidirectional and wide-field instrumentation: inexpensive all-sky cameras, microphones, and antennas with receivers for passive radar measurements and radio spectrum analysis.

### 6.2. *Overview of UAP observatory-class systems*

System development to date has focused on a proof-of-concept known as "Phase 1": the first observatory-class system in its temporary home on the roof of the Harvard-Smithsonian Center for Astrophysics (CfA). The instruments of the first observatory systems have been divided into five categories: (1) wide-field imagery for targeting aerial objects and triangulation of their position and kinematics; (2) narrow-field imagery and other observations which track objects across the sky; (3) radio receiver instrumentation for radar, radio spectrum analysis, and transponder message logging; (4) acoustic measurement devices (multi-band microphones); (5) environmental sensors for measuring parameters typically used to characterize terrestrial weather (pressure, temperature, humidity, wind velocity, and cloud cover) and space weather (local geomagnetic field and high-energy particles). In what follows, each of these categories is described along with a mention of relevant data processing approaches. The environmental sensors are given the most attention here because there is not a dedicated paper about these in this volume. Figure 2 contains a labeled illustration of the Phase 1 observatory-class system under development. Figure 3 shows photographs of the assembled instruments.

### 6.2.1. *Data archiving*

In Phase 1 development, as many of the data records will be archived as possible, in order to inform the development and refinement of our data processing pipeline and the development of future instrumentation for Phase 2. Also in Phase 1, we will experiment with ways of marking data for deletion in order to cope with the tremendous data acquisition volumes that will be typical of continuous operation (Cloete, 2023, this volume). Initially, time intervals marked for archival storage will be determined using just the wide-field and radar observations: i.e., in response to image- or radar-derived detection of objects in the local airspace. In future, this may be triggered instead by unusual readings in other sensors, especially if specific

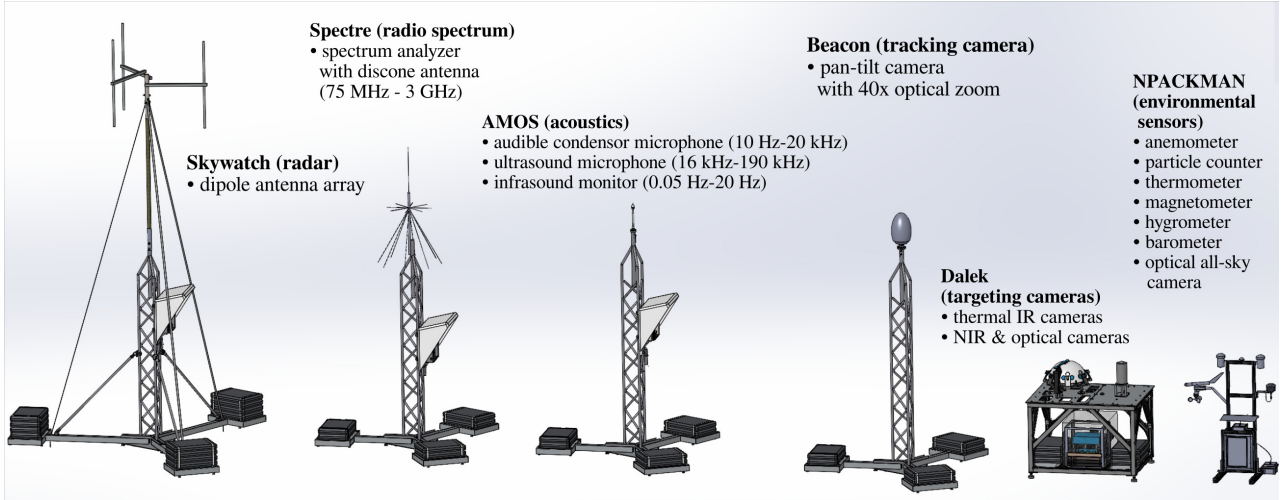

Fig. 2. Mechanical designs of the Phase 1 Observatory-class instrumentation suite on the rooftop of the Harvard-Smithsonian Center for Astrophysics (CfA).

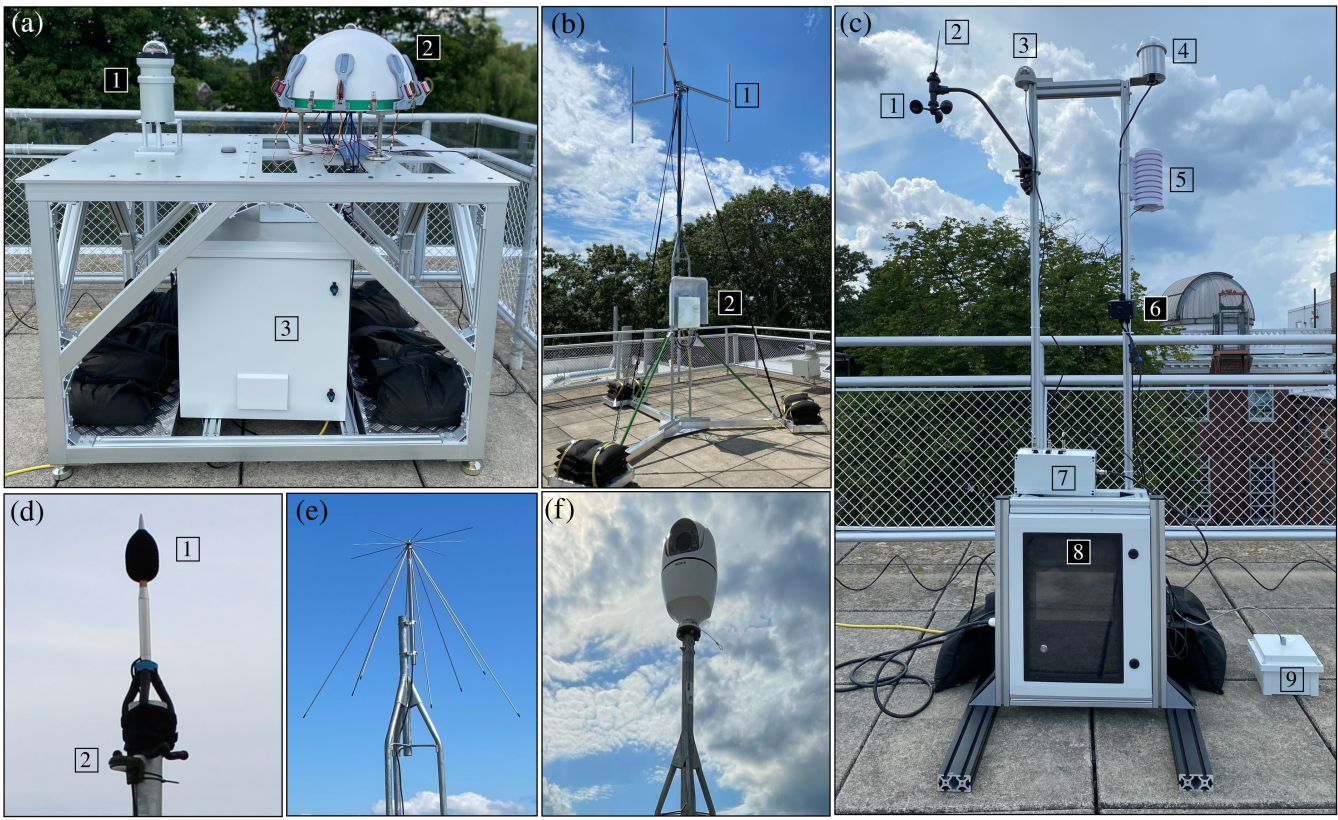

Fig. 3. Photographs of Phase 1 Observatory-class instrumentation suite on the rooftop of the CfA. (a) 1. ALCOR OMEA 9C all-sky camera; 2. Dalek hemispherical array of 8 LWIR microbolometer cameras (covered) and single optical + near-infrared all-sky camera; 3. Weatherproof enclosure for instrument computers. (b) SkyWatch radar tower: 1. dipole antenna array; 2. electronics enclosure. (c) NPACKMAN environmental monitoring system: 1. anemometer; 2. wind vane; 3. rain sensor; 4. optical all-sky camera; 5. temperature and pressure sensors; 6. camera fuse/switch box; 7. scintillation particle counter; 8. power and data module; 9. fluxgate magnetometer. (d) 1. GRAS audible microphone; 2. Wildlife Acoustics ultrasound microphone; infrasonic sensor not shown. (e) Discone wide-band antenna for spectrum analyzer (Spectre). (f) Beacon 8 narrow-field pan-tilt-zoom camera.

signatures in other sensors become associated with outlier events of special interest.

Ensuring the security, integrity, and authenticity of data is vital in a field where fraud has been frequently invoked for supposedly "extraordinary evidence." The protection of data records in storage and during transfer to servers in the cloud will be informed by the most recent Open Web Application Security Project (OWASP) guidelines regarding security risks and mitigation measures (OWASP, 2021). This includes (i) use of cryptographically strong algorithms for data storage such as the Advanced Encryption Standard (AES); (ii) storage of cryptographic keys using a key management service (KMS); (iii) hash-based integrity checks (checksums); (iv) use of multi-factor authentication for system access, and (v) ensuring software dependencies are up to date. The use of cryptographic signing is planned, to ensure the authenticity of data records.

The structure of our census database has yet to be determined in detail. Previous work has addressed the cataloging of UAP sighting reports and measurements, which may usefully inform our approach (e.g., Johnson & Saunders, 2002; Vallee, 2014).

### 6.2.2. *Wide-field observations*

These observations consist of wide-field imagery in optical and long-wavelength infrared (8 $\mu$m to 14 $\mu$m), acquired primarily for two purposes: (i) targeting the narrow-field observations (discussed in Sec. 6.2.3) and (ii) triangulating the positions of aerial objects and deriving their kinematics. All of the wide-field cameras will undergo an "intrinsic" calibration to determine the lens distortion function (a polynomial function of radial distance from the central axis) and an extrinsic calibration to compute the pointing direction matrix: i.e., the transformation from pixel coordinates to a vector in the 3-D Cartesian site frame (Szenher, 2023, this volume). The intrinsic calibration is accomplished for one camera of each type by staring at a grid pattern, followed by the application of image processing techniques for computing the distortion function. For optical cameras, the extrinsic calibration can be accomplished quasi-continuously under clear-sky conditions using astrometric plate solving (Szenher, 2023). Aforementioned ADS-B transponder broadcasts locate aircraft in the sky at precise times, which can be used to compute and update the calibration continuously, which is essential in the event of repeated minor camera displacements.

Wide-field observations can also be used to estimate apparent brightness of objects and in this way infer, in conjunction with range estimates, their intrinsic luminosity under conditions of perfect emissivity and low atmospheric absorption and scattering.

The wide-field instruments in the Phase 1 observatory class system consist of two all-sky cameras (the ALCOR OMEA 9C (30 MP) and ZWO ASI462 (2.1 MP)), and a set of eight thermal infrared cameras (FLIR Boson microbolometers) which provides coverage of nearly the entire sky (Szenher, 2023). Initial tests have shown that jet airliners ($\sim$60 m scale) are visible to a range of over 35 km in the Boson cameras in daylight. Light aircraft ($\sim$10 m scale) are visible to at least 7 km. These cameras are connected to low-power GPU-enabled single board computers for real-time processing with a machine-learning detection model that has been trained using a synthetic data set (Cloete, 2023, this volume). An additional custom-built 12 MP all-sky camera integrated with the NPACKMAN sensor system will be used to monitor the sky in the visible range; this will be supplemented with a camera sensitive to the mid UVB to UVA range in the near future. The angular coordinates of detected objects that remain in the scene for a sufficient time are transmitted to narrow-field pan-tilt-zoom (PTZ) instruments for tracking (Szenher, 2023). Also in real time, these objects will be linked across consecutive frames, producing tracklets that are later stitched together.

Observatory class systems will include one or more off-site cameras separated by at least $\sim$several km for triangulation. Angular coordinates will be communicated to the server at the primary site for the calculation of object position and kinematics. Objects for which there is no realistic solution (i.e., residing outside of a realistic detection volume) will be discarded, and assumed to consist of proximate clutter, such as insects and birds that appear in one camera only.

### 6.2.3. *Narrow-field observations*

The narrow-field observations in visible, near infrared, and thermal infrared require tracking of airborne objects to a precision of a few degrees to remain in the field of view. As mentioned, objects are identified for tracking if they persist in the wide-field camera field-of-view for a sufficient time. In the present system, narrow-field tracking is accomplished using cameras that physically rotate to capture objects and follow them across the sky. The only tracking instrument in Phase 1 is a pan-tilt Beacon 8 security camera (8 MP) with 40× optical zoom and roughly 2 arcsecond angular resolution (manufactured by Security Camera Warehouse (SCW)). For Phase 2, we are developing mirror-tracking systems that reflect light to stationary instruments such as narrow-field cameras, spectrometers, photometers, and polarimeters. The advantage of this latter approach is that lightweight mirrors can be rapidly swiveled with high precision and without excessive jitter. Secondarily, this allows light from the same optical path to be transmitted to multiple instruments. Mirror-tracking systems have been successfully used to track meteors with single arcsecond resolution at a rate of 100 frames per second (Vida *et al.*, 2021).

### 6.2.4. *Radio antennas and receivers*

Radio antennas and receivers are used for several markedly different purposes: (1) a multistatic passive radar system (SkyWatch; see Randall (2023) in this volume) for measuring object positions and kinematics; (2) a radio spectrum analyzer with omnidirectional wide-band antenna (Spectre) for measuring radio and microwave emissions; and (3) a whip antenna and receiver for logging ADS-B transponder messages to track known aircraft in the regional airspace.

The Skywatch multistatic passive radar system will be used to characterize the kinematics of objects measuring up to tens of meters in size throughout a region measuring up to 300 km across (Randall, 2023). The Phase 1 observatory instrument suite includes five passive antenna towers distributed on secondary sites throughout the region surrounding the primary observatory-class instrument site. Two of these are required for measuring a reference signal from regional FM radio stations (the reference receiver stations), and are located over 100 km from the primary site. The three-element array of dipole antennas on the remaining towers (the echo receiver stations) are used to record echoes from objects in the regional airspace.

The radar system operates in two modes: remote reference mode and direct reference mode. In remote reference mode, the reference transmitter signal is recorded at a receiver near the transmit antenna and is transferred via internet for correlation and subsequent analysis; this requires highly precise time and frequency synchronization. In direct reference mode, by contrast, the reference signal is also recorded by the echo receiver stations, and digital antenna null steering as well as adaptive noise canceling will be used to separately extract the echo and reference signals (Randall, 2023).

The time delay of echoes is used to infer total travel distance, which can be used along with transmitter and receiver positions to solve for aerial object positions. The correlation of echoes between receiver stations will be accomplished using a variety of approaches in order to compute and record the full set of possible solutions for a given event (Randall, 2023). The system will also be used to directly measure velocity from the Doppler shift of radar echoes which, in turn, can be used to calculate acceleration using finite difference. In addition, Doppler velocity dispersion may shed light on rotational kinematics. Skywatch is projected to be capable of measuring accelerations of up to $2800\,g \pm 4.5\,g$ and velocities up to $1.7\,\text{km/s} \pm 2.8\,\text{m/s}$. The detection range for an airliner is expected to be 150 km (20 dBsm); light aircraft at 110 km (15 dBsm), and birds at 6 km (−35 dBsm). Long-term development of the system will focus on miniaturization, so that lightweight and low-cost mesh systems can host smaller and cheaper antennas for very wide distribution and coverage. The Project may also acquire and evaluate active radar systems as part of Phase 2.

A second instrument called Spectre is a radio spectrum analyzer using an omnidirectional wideband discone antenna, and will be used to sample much of the VHF band (75 MHz to 300 MHz) and all of the UHF band (300 MHz to 3 GHz). An important role for this instrument is to monitor for the presence of signals used for remote-controlled aircraft such as drones. Secondarily, Spectre will be used to record unusual signals that may occur during UAP events, such as reported in McDonald (1972). The Project may add a capability for scanning much higher frequencies in Phase 2.

Finally, a tuned whip antenna with RTL-SDR receiver will be used to log ADS-B transponder broadcasts from aircraft in the regional airspace. These data will be used for the extrinsic calibration of wide-field cameras (Szenher, 2023) and for building acoustic and image training sets that can be used to train inference models.

### 6.2.5. *Acoustics*

The acoustic instrument suite (Audio Monitoring Omnidirectional System (AMOS)) has been described in Mead (2023) in this volume. These instruments cover a broad range of acoustic frequencies, and consist of the Infiltec INFRA20 infrasound microbarograph monitor (0.05–20 Hz), the GRAS Acoustics 41AC-3 CCP audible band condenser microphone (10 Hz–20 kHz), and the Wildlife Acoustics SM4BAT with SMM-U2 microphone ultrasonic condenser microphone (16–190 kHz). Using logged ADS-B transponder messages, we have generated a preliminary data set for identifying characteristic frequencies associated with conventional aircraft. The aerial census described in Sec. 3 will provide an enormous labeled database for training models to classify acoustic emissions from known vehicles. These data will also be used to characterize the variability in the sound path to help with identification and reconstruction of sound sources. Infrasound measurements will be compared to a regional network of similar sensors to identify signals derived from local versus distant sources.

The acoustic detection range for aircraft depends less on instrument sensitivity than it does on the ambient noise floor when using a single microphone. The approximate audible distance of an aircraft can be calculated using an accurate source level measurement. For example, the sound from a propeller aircraft (about 140 dBA at 1 m) would fall below the average ambient noise floor at approximately the following distances: 2.6 km "City" (56 dBA), 5.7 km "Suburban" (45 dBA), and 14.5 km "Rural" (26 dBA) (Mead, 2023).

### 6.2.6. *Environmental sensors*

The environmental sensor package will consist of a thermometer, barometer, hygrometer, rain gauge, and anemometer to measure ambient atmospheric conditions, as well as a muon detector, a magnetometer, and a wide-field camera already described in Sec. 6.2.2. For the Phase 1 observatory, all of these instruments reside in the NPACKMAN environmental instrument system. Because there is no paper in this volume dedicated to describing these instruments, we provide extra discussion here of the justification and technical specifications.

*NPACKMAN:* The New PArticle Counter k-index Magnetic ANomaly (NPACKMAN) is a unique instrument developed for two goals: (i) to characterize the environment, providing measurements of meteorological variables; and (ii) to investigate the impact of space weather on the Earth's surface and near subsurface. NPACKMAN is an upgrade from PACKMAN (Mathanlal *et al.*, 2021) and was developed specifically for the Galileo Project. NPACKMAN makes use of more sophisticated sensors and electronics than its predecessor to improve the accuracy of data and scientific return. These sensors have been revamped using certified, industrial grade components, including the use of water resistant enclosures. The circuitry is galvanically isolated and care has been taken to ensure the electromagnetic compatibility of components. These steps ensure that the data generated from the sensors are of highest quality, without influence of noise or cross-contamination between different sensors. The quantities measured by NPACKMAN are summarized in Table 7.

The core of the instrument package consists of a power and data module that houses the industrial grade computer, communication module, and power conditioning circuitry. Sensor modules interface with the power/data module through weatherproof connectors and the entire system is IP65 rated for ingress protection according to International Electrotechnical Commission (EN/IEC) 60529 standards. It is modular and can be extended with more sensor modules should the need arise.

Table 7.  Measured quantities and specifications of NPACKMAN sensors (PE = pending evaluation).

| Quantity | Range | Accuracy | Resolution | Sampling rate |
|---|---|---|---|---|
| Temperature | $-40°$ C to $+85°$C | $\pm 1°$ C | $0.01°$C | 1 Hz |
| Relative humidity | 0%–100% | $\pm 3$% | 0.04% | 1 Hz |
| Pressure | 10–2000 mbar | $\pm 2$ mbar | 0.016 mbar | 1 Hz |
| Precipitation | 250 mm/hr | $\pm 10$% | 0.02 mm | 1 Hz |
| Wind direction | $0°$ to $360°$ | $\pm 3°$ | $22.5°$ | 0.45 Hz (on average) |
| Wind speed | 1 to 322 km/h | $\pm 5$% | 1 km/h | 0.45 Hz (on average) |
| Magnetic field | $\pm 50\,\mu$T | PE | 1 nT | 1 Hz |
| Muon coincidence detector | 0.3–5 GeV (muons); 100 keV to 5 MeV (other particles) | PE | PE | 1 Hz; $\frac{1}{60}$ Hz (coincidence) |

*Particle counters*: The Galileo Project will make use of two kinds of energetic particle counters: Geiger–Müller counters rely on ionization of a gas confined in a tube at high voltage, and scintillation counters rely on scintillating materials that emit light as high-energy particles pass through. The Phase 1 observatories will include just a scintillation counter in NPACKMAN, which is sensitive to gamma radiation in the range 100 keV to 5 MeV, and can detect muons, fast neutrons, and charged subatomic particles. Based on the photomultiplier signal amplitude, this detector will be able to estimate particle energies. Operating in a coincidence mode, which leverages a stacked pair of scintillators and photomultipliers, this sensor is also able to calculate the fraction of detections that derive from cosmic ray muons whose average energy is 4 GeV at sea level.

*Magnetometers*: Magnetometers are commercially available for a wide range of price and performance capabilities, depending on the type of magnetic sensor they use (e.g., fluxgate, Hall effect, magnetoresistive) as well as electrical shielding and temperature compensation. A magnetometer's specifications include the measurement range, noise level, thermal drift, accuracy, and resolution.

When properly encased and buried underground, the Mag-13MS100 fluxgate magnetometer by Bartington meets the requirements outlined in Sec. 5 and has been acquired for the Phase 1 observatory. The magnetometer in NPACKMAN satisfies most of these conditions, but its limited dynamic range ($\pm 50\,\mu$T) is appropriate for monitoring geomagnetic variations and will saturate in the presence of a powerful field. The collected data will be analyzed concurrently with the publicly available magnetic field data from nearby INTERMAGNET observatories (Love & Chulliat, 2013): e.g., the Fredericksburg (FRD) and the Ottawa (OTT) magnetic observatories, operated by the U.S. Geological survey and the Geological survey of Canada, respectively.

## 7. Site Selection

As described in Secs. 3 and 5, the project's principal scientific objective is to perform a comprehensive census of aerial phenomena, which will be carried out in diverse locations and under diverse conditions. Meeting this objective requires that we develop the instrumentation system, which in turn requires one or more locations that are suitable for long-term instrument research and development. A second requirement is to conduct the census itself: i.e., to characterize aerial phenomena in U.S. airspace using a sample of up to 10 regions and a deployment of up to 5 years in the initial project phases (subject to future funding constraints and opportunities). This requires relatively autonomous facilities at locations that are representative of diverse geographies, atmospheric and climate conditions, population densities, and air traffic. Apart from the observatory system located on the rooftop of the CfA, most or all of the other sites will not be disclosed until the census data are published, in order to protect the equipment and the privacy of site proprietors. We anticipate there may be hoax attempts; we have designed the procedure outlined in Sec. 3 to classify such events as "nominal." We will conduct our own data injection tests and spoofing events to rigorously test our system.

### 7.1. *Definitions*

The observatory-class systems described in Sec. 6.2 require a primary site surrounded by multiple, smaller secondary sites. Secondary sites host ancillary instruments that capture critical information for triangulation of aerial phenomena, to derive their range and kinematics, as well as corroborating sensor data. One set of the secondary sites will host individual radio antennas, two of which are located up to 150 km from the primary installation and three of which are within 30 km (see Randall (2023), this volume). Two more secondary sites will host individual cameras located within several km of the primary installation. Secondary sites may also host acoustic and magnetic sensors in order to distinguish local from regional variations in both the signal and the background.

The site selection process will occur on several geographic scales; we will use the following terminology to identify these:

- *Region*: A local name for a large area, usually with characteristic geographical features, that may be a town name, and/or encompass multiple towns or counties, e.g., "Boston," which refers to the greater Boston metro area.
- *Location*: A fixed-size geographic area centered on a primary, multi-instrument site, and including all secondary sites (of order 100 km in diameter).

- *Primary site*: The host for the main, multi-instrument suite; its dimensions are roughly 20 m × 20 m.
- *Secondary site*: The host for either a radio receiving antenna or all-sky camera, along with acoustic, magnetic, and other corroborating instrumentation; its dimensions are roughly 6 m × 6 m.

## 7.2. Site requirements for instrument development

The task of using ground-based instrumentation to identify every known object in the sky and carefully characterize unknowns is formidable. This requires the seamless fusion of the multiple measurement modalities described in Sec. 6.2, combined with a data analysis pipeline and in-depth understanding of ambient environmental variables. The work as a whole encompasses multiple fields of research, from Earth and planetary science to audio, optical, radio, computer, and mechanical engineering. The rapid prototyping, testing, and refinement cycle requires a stable long-term development site that is easily accessible to the technicians, engineers, and scientists involved in the work. We envision the following site categories and stages for the development and testing of observatory systems.

### 7.2.1. Phase 1 proof-of-concept site

Our instrument development starts with Phase 1, a proof-of-concept study based on the roof of the CfA, which rises 27 m above sea level and is located 10 km west of Boston Logan International airport. During daylight hours, aircraft continuously depart from and land at Logan airport on a path that crosses approximately 2 km to the northwest at an elevation of 1500 m. Jets routinely pass overhead at altitudes from 7 to 12 km on trans-Atlantic routes. Helicopters and propeller aircraft from nearby smaller airports also frequently pass into close proximity of the CfA. This site hosts the instruments described in Sec. 6.2.

The Phase 1 observatory also comprises five secondary sites that each host a single radio receiving antenna. Two of these sites are located near FM radio transmission towers to measure reference signals. The three other antenna sites are located in the greater Boston area approximately 25 km from the primary site at the CfA. These six antennas are part of the experimental passive radar system developed to detect, localize, track, and characterize the kinematics of moving objects in the sky described in Sec. 6.2 and Randall (2023) in this volume.

*Criteria for the Phase 1 site*: of necessity, the primary site was located on Harvard-owned property with sufficient outdoor space, an indoor staging lab, with easy access by nearby personnel, along with access to power, internet, security, and air traffic recorded using ADS-B receivers for model training and instrument testing. The secondary radar sites have the same basic requirements: size, power, internet, security, but must also comply with local ordinances and, importantly, must have a generous and welcoming host. Two of the radar sites require proximity to FM transmitters within 150 km of the CfA, and three require a radial distance within 30 km of the CfA.

### 7.2.2. Development site

Once the instruments are assembled, are operational, and have successfully completed initial testing, the primary site will be moved from the CfA roof and relocated at a long-term development site. At this site, the instrument suite and analysis software will be tested, trained, calibrated, and refined until it is ready for duplication and deployment to the first testing sites. Here also, future Phase 2 instruments will be developed, tested, and refined. The existing radar sites from Phase 1 will remain in place, and two additional secondary sites for optical triangulation within several km of this primary site will be secured.

*Criteria for the development site*: The primary site should be located within the greater Boston area in order to take advantage of the existing secondary radar sites. The property must have sufficient space, an indoor staging lab, access to power, internet, security, air traffic, and proximity to multiple field engineers and scientists involved in the Project.

### 7.2.3. Testing sites

The Project intends to duplicate the Phase 1 instrument suite and establish one or two observatory systems at testing sites for long-term field testing in areas of potential interest. As such, systems at testing sites are intermediate between those at development sites and deployment sites. That is, they are located in environments different from the development site but must be accessible to Project field engineers who can install, calibrate, troubleshoot, and stabilize the hardware and software in the new environments. The testing sites are not likely to be situated on Harvard-owned property, and therefore require welcoming individuals or institutions as hosts. These sites are selected primarily for logistical reasons and secondarily for optimization of viewing and detection of aerial phenomena. The information gained from working in these testing sites feeds back to the development site and will inform Phase 2 development. As Phase 2 equipment development is completed, these instruments will be duplicated and sent to the testing sites for testing, calibration, stabilization, and data collection.

*Criteria for testing sites*: Each site requires a host property within an hour of at least one Project field engineer, sufficient outdoor space, access to power, internet, security, air traffic. Also important is the ability to secure four or five radar and two all-sky camera secondary sites, which may host ancillary acoustic and magnetic sensors. High visibility of the sky and horizon is important for testing sites. In the best case, the recent history of UAP reports or reporting statistics in the surrounding region should suggest a higher-than-average likelihood of future UAP sightings (see Sec. 7.3.3).

## 7.3. Site requirements for long-term census data collection

Our objective is to conduct a census of aerial phenomena, as described in Secs. 3 and 5. The census will reveal a detailed distribution of aerial phenomena through time (up to ∼5 years), but the small number of observatory-class systems currently planned (up to ∼ 10) will permit at best a limited characterization of the distribution over the continental U.S. (The number of stations and duration of data collection is subject to future funding constraints and opportunities.) Long-term census data collection places particular conditions on deployment sites, including ones related to experimental design, hypothesis testing, and building a representative census of aerial phenomena.

After Phase 2 instrumentation is completed and is successfully collecting data at testing locations, multiple copies of the instrument suite will be assembled for the remaining seven deployment locations. Ideally, these are "turn-key" systems, requiring infrequent visits from a Project field engineer, and are also modular, allowing easy upgrades of swappable components.

The basic logistical requirements for deployment sites are expected to be similar to testing sites, but without the need for such close proximity to a field engineer. Air traffic will continue to be helpful, to allow for continuous calibration with ADS-B-derived identifications of overflights. It is anticipated that solar-powered battery and stand-alone 5G hotspots will replace the need for some sites to have power and internet. In addition, the instrument requirements of Phase 2 may further constrain deployment site selection in ways we cannot predict at this time.

### 7.3.1. Siting considerations based on logistics

Any long-term deployment site must meet the set of basic logistical requirements outlined above. Willing and generous individuals or institutions must be identified who will host the equipment in a secure environment for a period of up to five years. Each site at the location must have sufficient space, access to power and internet, be accessible by truck for the installation of equipment, and be readily accessible and able to accommodate visits from Project field engineers. Each deployment requires a primary site and seven secondary sites. As such, a single deployment may involve multiple property owners.

Once a region is selected, a process is initiated to identify hosts and select a specific deployment location encompassing the eight sites. For each individual site owner, this process begins with an initial engagement by a Project member. If the owner is willing to host, then community engagement begins: a process that includes understanding local bylaws and ordinances, developing positive relations and identifying safety or security concerns, and initiating local Indigenous relations and tribal land engagement. Finally, site-specific legal requirements are drafted, which cover agreements, liability, and insurance. Once these are signed, the logistical challenge of shipping, storing, and staging the deployment is engaged. Deployment itself involves on-site assembly, installation, validation, and testing. In addition, the site must be secured against vandalism and accidents. Steps should be taken to reduce any outstanding liabilities for the landowner or the Project. Once the site is secured and operating, there will be ongoing site monitoring and maintenance responsibilities, some of which may be addressed by the host, if appropriate. When the experiment is completed, a Project engineering team will return for tear-down, site restoration, and return shipping of equipment for storage at a Harvard facility.

### 7.3.2. *Siting considerations based on environmental conditions and effective detection volume*

A reasonable approach to site selection can begin with maximizing the total volume of sky sampled over the course of the experiment. For each site, the volume of sky observed depends both on the instrumentation and fixed obstructions such as hilly terrain, mountains, nearby buildings, and tree cover, as well as on time-dependent conditions such as weather and atmospheric visibility. Visibility may in turn depend on geography and the natural environment. For example, coastal areas in southern California may have fewer days with fog than those in the coastal northeast of the U.S. Alternatively, visibility may be related to population. For example, an area near a national park in Montana may have clear air, and thus higher visibility, while a densely-populated city like Philadelphia may sit under a summer haze, with relatively significant air and light pollution. Selecting 10 deployment locations where sky viewing is optimal would bias our sample to specific terrains, ecosystems, weather patterns, and levels of urbanization.

While sky visibility and the optical detection volume are the most important factors in this category, consideration of factors affecting other instrument detection volumes is also important, such as the ambient acoustic noise floor.

### 7.3.3. *Siting considerations based on frequency of UAP reports*

Networks of scientific instruments are expensive to build, maintain, and transport. Maximizing the value of this investment requires maximizing the probability of exposure to potential UAP, at least for a subset of the deployment sites. The approach described here develops a probability model of UAP reports, based on a large existing database of historical UAP reports. This model is used to suggest regions where UAP sightings may be expected to occur with greater frequency.

There are special challenges associated with locating areas of elevated concentrations of UAP reports using existing databases. Chief among these is the strong correlation between reports and population — the geographical distribution of events (Fig. 4 left) almost exactly mirrors the distribution of population (Fig. 4 right). Thus, the U.S. counties with the most reports are the counties with the largest populations. At the same time, calculating a figure like "reports per capita" is not much more informative. Locales with high reports per capita are often small towns with few events, i.e., implying an annual event rate of nearly zero.

Instead, we take an outlier detection approach: we develop a model of the frequency of reports, and compare each locales' expected reports with observed reports. Population density and land area prove to be sufficient independent variables to capture most of the variation and generate reasonable estimates. Model errors — the difference between the measurements and model predictions — are interpreted as "excess" sightings, over and above what would be expected for the territory. Territories with the largest excess can thus be considered outliers and for this reason noteworthy.

The sightings report database used for this study was collected by the National UFO Reporting Center (NUFORC, Davenport (2022)). This civilian organization has collected UAP reports from the public since 1974. Initially collected via a telephone hotline, they are now submitted online and uploaded to a public website. Staff occasionally annotate reports, marking obvious sightings of satellites and hoaxes. A typical entry includes an eyewitness description of the event, the date and time, location, shape of observed phenomenon, and the duration of the sighting. The dataset for this study was scraped from the NUFORC website on May 26, 2021 (Davenport, 2022). The reports were processed using a series of automatic cleaning and standardization techniques, to ensure that spellings are accurate and consistent, and to assign each municipality to its county and a latitude and longitude; these were then compiled into a central database. Data were filtered to include only the continental United States between 2010 and 2020. A total of 48,531 reports were included in the analysis. Nearly ten thousand cities and towns are included, encompassing all large- and medium-sized cities.

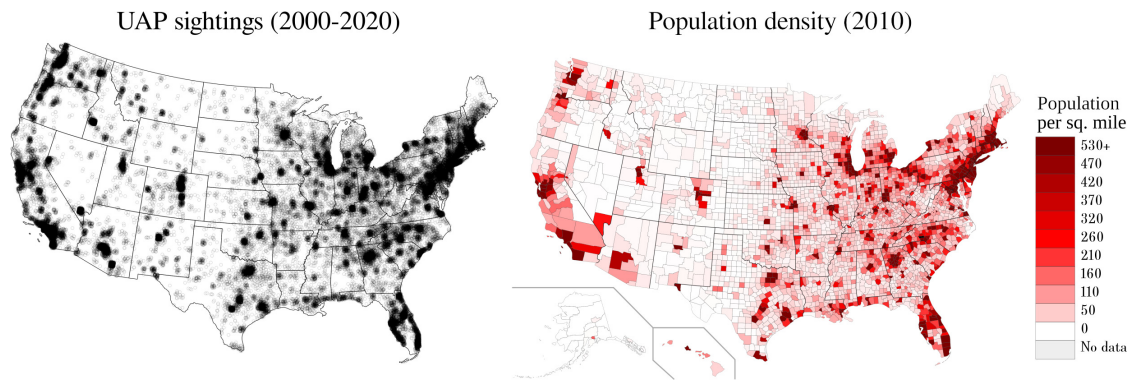

Fig. 4. Left: UAP sightings in the continental United States (2000–2020; NUFORC (Davenport, 2022)); each circle represents one sighting. Right: Population density (U.S. Census Bureau, 2011).

Data for the explanatory variables, population and land area, were retrieved from the U.S. Census Bureau (U.S. Census Bureau, 2011). The best and most granular data was available at the county level. This necessitated mapping the NUFORC counts from city- to county-level. The 2020 U.S. Census provides yearly population and population density from 2010 to 2020 for counties. The land area is available as of 2010 (U.S. Census Bureau, 2011).

The association between a county's number of sightings and the independent variables is estimated using mixed effects linear regression (Gelman & Hill, 2006; Galecki & Burzykowski, 2013). For each individual county $i$ in year $j$, the number of UAP reports $y_{ij}$ is modeled using regression, with land areas $x_{1i}$, and population density as $x_{2ij}$. Because of the longitudinal nature of the data, random slope effects are included for county $i$ in year $j$ (i.e., $\alpha_i$ and $\gamma_j$, respectively). Both are normally distributed around zero with a standard deviation $\sigma_\alpha$ and $\sigma_\gamma$, respectively:

$$y_{ij} = \mu + \beta_1 x_{1i} + \beta_2 x_{2ij} + \beta_3 x_{1i} x_{2ij} + \alpha_i + \gamma_j + \epsilon_i,$$

where

- $\mu$ is the intercept term, representing overall mean number of sightings
- $\alpha_i \sim N(0, \sigma_\alpha)$ is the 'effect' of the county $i$
- $\gamma_j \sim N(0, \sigma_\gamma)$ is the 'effect' of the year $j$
- $\epsilon_i \sim N(0, 1)$ is the random error

The dependent variable was logged (base 10 logarithm), whereas the two independent variables were logged and normalized (omitted in the notation above for clarity). Additionally, an intercept term is included to test the possibility that the effect of population density varies depending on land area, and vice versa. Parameters are estimated to maximize the log-likelihood of the data, using R's lmer4 package (Bates *et al.*, 2015).

The parameter estimates are given in Table 8. All terms are highly significant ($p < 0.01$). The plot in Fig. 5 is a graphical visualization of the model. It demonstrates how the number of annual UAP

Table 8. Estimated model parameters and model performance metrics.

| | |
|---|---|
| Model parameters: | |
| Land area ($\beta_1$) | 0.558*** |
| | (0.012) |
| Population density ($\beta_2$) | 0.481*** |
| | (0.007) |
| Interaction term ($\beta_3$) | −0.027*** |
| | (0.004) |
| Constant ($\mu$) | 0.783*** |
| | (0.048) |
| Performance metrics: | |
| Total counties × years | 11,785 |
| Log Likelihood | −9,823.979 |
| Akaike Inf. Crit. | 19,661.960 |
| Bayesian Inf. Crit. | 19,713.580 |
| Conditional $R^2$ | 0.673 |
| Marginal $R^2$ | 0.484 |

*Note:* * $p < 0.1$; ** $p < 0.05$; *** $p < 0.01$.

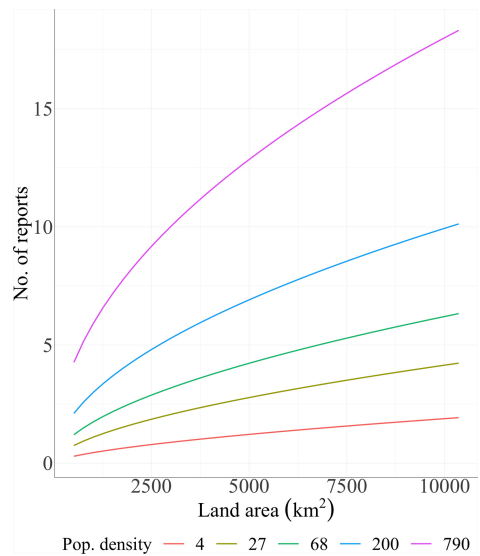

Fig. 5. Graphical representation of the fit model: number of reports per year vs. population density in km$^{-2}$.

Table 9. Locations with the 10 largest report excesses (interpreted as outliers).

| State | County | Largest cities | Reports | Prediction | Excess |
|-------|--------|----------------|---------|------------|--------|
| Arizona | Maricopa | Phoenix; Mesa | 1137 | 784 | 353 |
| California | Los Angeles | LA; Long Beach | 947 | 734 | 213 |
| Washington | King | Seattle; Auburn | 681 | 488 | 193 |
| South Carolina | Horry | Myrtle Beach; N. Myrtle Beach | 385 | 229 | 156 |
| California | San Diego | San Diego; Escondido | 592 | 454 | 138 |
| California | Orange | Anaheim; Huntington Beach | 491 | 366 | 125 |
| Illinois | Cook | Chicago; Oak Lawn | 522 | 416 | 106 |
| California | Riverside | Riverside; Temecula | 485 | 380 | 105 |
| Nevada | Clark | Las Vegas; Henderson | 442 | 348 | 94 |
| Oregon | Multnomah | Portland; Gresham | 317 | 226 | 91 |

sightings in a "generic" county varies according to land area and population density.

Table 9 shows the counties with the largest model errors over the decade under study. This list is substantially different from both a list of the highest absolute number of reports and a list of highest per capita reports. It is these locations that could merit further examination as potential "hot spots" of UAP activity. As noted, the excess reports in these regions may instead reflect other factors such as visibility (annual weather conditions), the presence of airports and military bases, social and cultural differences, as well as media interest and the presence of enthusiast organizations.

As usual, regression analysis identifies statistical associations only, and makes no claim about causality. It identifies areas with an unusually high number of NUFORC reports. These reports are "filtered" through the humans that make them, and the relationship between such reports and the overall UAP phenomenon (observed *and* unobserved) is not assumed. This analysis does not explain *why* these areas are outliers; nor is it guaranteed that sighted objects are truly anomalous. Variables such as local climate and average sky visibility, social and cultural factors (including media attention and activities of local interest groups), nearby military bases, etc., should be considered in follow-up analyses.

The relationship between locations derived from an analysis such as the one described above (based on historical UAP reports) and the locations of future manifestations of observational anomalies is unknown. However, the aerial census data can be used to determine whether contemporaneous UAP reports filed with organizations like NUFORC can be classified into categories of known phenomena. As mentioned at the start of this work, we are not primarily concerned with evaluating the veracity of UAP reports. Nevertheless, this kind of comparison may assist with using UAP reports to decide where to locate future experiments. Researchers in the future may also use this kind of comparison to evaluate more accurately the information content of historical UAP reports.

It should be noted that regions of interest for long-term observation (ROIs), identified on the basis of the statistics of historical reports, will sometimes not overlap significantly with ROIs based on the optimization of total effective detection volume. For example, although historical UAP reports strongly correlate with population, areas with high population may have lower sky visibility.

Instead of relying on reports posted by members of the public on UAP reporting websites, where the accuracy and quality of reporting varies enormously, ROIs can be chosen on the basis of reporting by expert witnesses in high-profile events. For example, eyewitness testimony from U.S. Navy pilots and other U.S. Navy personnel since 2004 can form the basis for highlighting two regions reported to have ongoing sightings. These are the Catalina Island region in California located near the Nimitz incident 200 km to the south (Powell *et al.*, 2019), and the mid-Atlantic coastal region near the naval training airspace 16 km off Virginia Beach, VA (Cooper *et al.*, 2019). Alternatively, a few regions of interest may be selected because of a relatively continuous level of elevated UAP activity, as documented by researchers who have studied eyewitness accounts or who have conducted instrument-based field investigations (Teodorani, 2004) and published in the peer-reviewed or gray literature.

### 7.4. Summary of procedure for site selection

The identification of instrument sites will begin with meeting a set of logistical requirements. Added to this are the constraining environmental factors, such as viewing conditions, terrain, and noise floor. Next, consideration is given to criteria based on historical UAP reports, where some of these may derive from a statistical analysis. Finally, site selection can be informed by hypotheses generated from previous data collection. In summary, the categories of site criteria are these:

(1) *Logistical criteria*: e.g., welcoming hosts, power, internet accessibility, security, proximity of field engineers

(2) *Environmental criteria*: e.g., sky visibility, favorable atmospheric conditions

(3) *Criteria derived from historical UAP reports*: e.g., statistical excess of reports

(4) *Criteria for hypothesis testing*: e.g., placing instruments in specific regions to compare aerial phenomena detection rates against UAP reporting based on specific factors or a predictive statistical model.

The selection of development sites depends almost entirely on logistical constraints, such as finding a host property close to Project field engineers; it depends somewhat on environmental factors, with a preference for rural areas. The testing site selection will be informed by similar logistical considerations, but environmental factors and historical UAP reports will also play a role. Deployment site selection will depend on only the most basic logistical considerations, such as the need for hosting, and will not require proximity to field engineers. For deployment sites, the environmental factors and historical UAP criteria will play a much larger role. The top-ranked deployment sites will be subject to further logistical requirements that may have arisen during the course of instrument development cycles. Also, specifications for creating test and control sites may be applied, based on hypotheses derived from lessons learned during the development and testing stages of the project.

## 8. Conclusions

UAP present a long-standing mystery that can and should be investigated by the tools of contemporary science. In contrast to most previous work, the UAP branch of the Galileo Project will focus on collecting new data using calibrated instruments that are entirely controlled by a team of scientific investigators, applying established techniques for ensuring the authenticity and integrity of data records, and publishing data and results in peer-reviewed journals. To this end, the project is developing an integrated multimodal system of instruments and software for parallel and simultaneous detection and characterization of aerial phenomena. We have supplied a detailed science traceability matrix (STM) to connect project goals and objectives to project-level requirements, sought-after physical parameters, relevant observables, and instrument measurement requirements. The Galileo Project observatories will be used to conduct a census of aerial phenomena in a wide range of locations and under a diverse set of conditions. Statistical and machine learning techniques will be used to sort these into categories of known phenomena, while also marking artifacts, ambiguous detections, and statistically distinctive outliers. Outliers will be investigated and evaluated through targeted follow-up observations as potential statistical anomalies: as phenomena potentially unknown to current science. A process of hypothesis formulation and testing will follow in order to determine whether statistical anomalies fall outside of expectations based on prevailing scientific beliefs (scientific anomalies). Sites for locating the Galileo Project observatories will involve consideration of specific criteria required for the development, testing, and deployment of our instruments.

## Acknowledgments

WW thanks mentor and friend David Akers for many stimulating conversations. Christopher Mellon, Jacques Vallée, Nick Pope, Florin Stefan-Morar, and Robert Powell are warmly thanked for their helpful reviews and insights. We also thank Jacob Haqq-Misra for his insights and contributions. We are grateful to Hassaan Tariq for assistance with citations, as well as Timothy Tavarez and Bradley Reimers for their contributions to software development. Lily Kuan is thanked for assistance with instrument deployments.

## Appendix A. Common Objections to the Study of UAP

While this work is focused on positive motivations for investigating UAP (Sec. 2), in what follows we address major misconceptions and commonly-occurring objections to the study of UAP. Many of these have been gathered from other sources, including those addressed in Wendt & Duvall (2008) and briefly in Hynek (1966).

(1) *UAP characteristics are impossible.* The suggestion in this case is that behaviors and properties ascribed to UAP through eyewitness reports and derived from limited sensor data are *prima facie* impossible. It follows from this that no study should be undertaken because there is no reason to investigate the evidence for impossible phenomena. In addition, UAP researchers are frequently reminded that extraordinary claims require extraordinary evidence.

It is true that apparently impossible characteristics have been found in many cases to have plausible prosaic explanations. For example, optical mirages, reflections, and image splitting can in some cases account for some of the unusual characteristics ascribed to UAP, such as supersonic speeds without noise, or extreme accelerations. It is important to emphasize that these known phenomena do not currently suffice as a general scientific explanation for many UAP reports. For this, a rigorous and systematic program of scientific inquiry is required. The behaviors ascribed to UAP have rarely been sufficiently characterized using scientific instrumentation to assess impossibility, even with respect to known laws of nature and the capabilities of modern technology.

Unfortunately, the history of science is littered with bold claims about technological feats that are impossible, and which later come to pass. Heavier-than-air flight and space travel are two prominent examples of this failure of imagination (Dalamagkidis *et al.*, 2012). What counts as 'impossible' is determined with respect to both prevailing scientifically-informed theories and current scientific dogmas. It has happened many times throughout the history of science that supposedly impossible events and evidence were consistently denied or ignored until a new set of beliefs was able to accommodate them. Examples include the resistance to continental drift in the absence of a plausible mechanism (Oreskes & Grand, 2003), and to the earlier-mentioned historical record of meteorite falls before the discovery of asteroids (Marvin, 2006).

In short, if extraordinary claims require extraordinary evidence, scientific researchers should be encouraged to investigate such claims by collecting evidence in a systematic and unbiased fashion. That is, extraordinary claims with intriguing but subpar evidence should be investigated rather than ignored.

(2) *Extraterrestrials cannot reach us* and, if they could, *they would land on the White House lawn.* We draw no equivalence between UAP and extraterrestrial craft and we begin this work with an agnostic position regarding the ultimate nature, provenance, and intent (if any) of these phenomena. Our attitude is consistent with the public agnosticism described by Wendt & Duvall (2008), who have urged suspension of judgment and a willingness to investigate.

The implicit bias equating UAP with extraterrestrial craft derives partly from a ubiquitous association in the popular culture, as well as claims that UAP can accelerate to planetary escape velocity (Knuth *et al.*, 2019). Also relevant is the often-cited expression "any sufficiently sophisticated technology is indistinguishable from magic" (Clarke, 1984): i.e., seemingly magical events may imply an extremely sophisticated nonhuman technology. Many scientists believe extraterrestrial contact is highly unlikely owing to the low odds of long-lived extraterrestrial civilizations arising in the first place, or the vast distances that must be traversed to reach Earth from even the nearest star system. By contrast, many astronomers routinely teach their introductory students about the Fermi Paradox, according to which extraterrestrial visitation might be *expected* under certain conditions. That extraterrestrials would "land on the White House lawn" is remarkable for how much it assumes about the intentions of entities about whom presumably nothing is known in advance.

(3) *We would know by now if this were real.* The latest version of this objection relies on noting the prevalence of camera-enabled smartphones: i.e., with so many instruments in the wild, if this phenomenon were in fact real then it should be a well-documented and well-established fact by now.

Photographs acquired using hand-held devices and videos of purported UAP, which are prevalent on internet video posting sites, overwhelmingly fail to meet the evidentiary standard required for scientific study. This concern relates to the primary motivation for our work, which is to collect scientifically rigorous, multimodal measurements using calibrated instruments to establish whether UAP represent a class of phenomena currently unknown to science.

Material posted on the internet is usually without essential context and without a well-documented provenance and chain of custody, and many of the most remarkable images and videos could be easily hoaxed. After accounting for optical reflections and artifacts from windows and lenses, imaging sensors, and software processing, video and photographic evidence is only as credible as the witness who reports this evidence, highlighting the importance of ensuring that sources are authentic. The Galileo Project will address this problem by collecting secured data as part of a years-long scientific experiment involving well-characterized scientific instrumentation that is multimodal and multispectral, permitting corroborative detection and characterization of outlier events.

The fact that many UAP incidents have been hoaxes does not imply that all of them are hoaxes or that the phenomenon lacks objective reality. The history of science has seen its share of charlatans and frauds regarding claims about phenomena that were eventually borne out with verifiable evidence. The case of the Piltdown Man hoax is a good example: a "missing link" homonid skull was derived from the bones of a human and an ape (Straus, 1954). In spite of this hoax, plenty of verifiable fossil evidence for early homonids was eventually discovered and this prediction of Darwinian evolution was confirmed.

The expectation that there should be many convincing eyewitness recordings implies multiple challenges that are worth noting as well. That is, whenever someone observes something that is unprecedented in their experience they may not, as a rule, reach for their camera. This is partly because the accurate perception of something that is so completely novel is time-consuming, and can be affected by confusion and fear (Haines, 1980). In light of this, it is unsurprising that many witnesses focus on absorbing as much detail as possible using their naked senses. This is not to mention the challenges connected with capturing a focused high-resolution image using a smartphone camera in a matter of seconds

(e.g., glare on the screen, unsteady hands, and "aiming" a flat object).

A more interesting question concerns meteor detection systems, which have been in operation for decades. Archives of these data should be examined for candidate anomalies. Unfortunately, some of these systems automatically discard images that do not contain linear streaks produced by meteors at high altitude (Jenniskens *et al.*, 2011). That is, at least some of these systems are designed to capture and record only the phenomenon that is sought (meteors). It is also undoubtedly true that the defense establishment has used very sophisticated multi-sensor systems to monitor the national airspace and have collected a vast and highly relevant data set. Unfortunately, we have access neither to the instrument specifications nor the data they have collected.

Finally, it is worth recalling that UAP reports, which share many characteristics in common, have continued despite repeated assertions that these phenomena must be a transient popular delusion. Harvard astronomer Donald Menzel was incorrect in his 1972 forecast: "I do predict... a continued decline of public interest in UFO's" (Menzel, 1972) and "scientists of the twenty-first century will look back on UFO's as the greatest nonsense of the twentieth century." The fact that the phenomenon has persisted for so long suggests that it is not a transient craze, that it may indeed have objective reality, and that the scientific community should at least investigate to find out.

(4) *Scientists shouldn't "sell out" to popular interests.* This work will address a persistent phenomenon that happens to be of interest to some fraction of the public and which has wide representation in the popular culture. Science can gain public confidence and support by demonstrating how a scientific approach can be used to rigorously address problems of popular concern that have been ignored for decades. Moreover, as noted in Page (1972), the UAP topic provides an excellent chance to educate the public about science in action, critical thinking, and about the technologies and natural phenomena to which this topic naturally connects.

(5) *The study of UAP is inherently pseudoscientific.* This objection commonly relies on concerns

about the study of UAP reports, which more closely resembles detective work than it does research in the "hard sciences." UAP studies may be regarded by some as inherently unscientific because they assume there is no relevant "hard data": i.e., no quantitative data collected using calibrated instruments. A detailed inspection of the UAP literature demonstrates that this is not obviously true. But even conceding the point, this provides the rationale for undertaking a prospective study of this kind, which aims to acquire hard data using a rigorous experimental design.

A second inspiration for this line of argument is that such work is attempting to prove the existence of something that it assumes to be real, and that science is not concerned with such endeavors. This argument would invite us to forget decades-long efforts to demonstrate, for example, the existence of gravitational waves, since such searches are powerfully motivated by the expectations of theory. To be clear, the present study sets out to investigate *whether there are* anomalies that can be scientifically verified, rather than prove the existence of, for example, flying saucers that some may ardently wish to be real.

Finally, the allegation of pseudoscience may stem from observing that the problem this work attempts to address is not "well-posed." This may derive from a conventional wisdom that the only problems worth pursuing are the kinds that fit the "puzzle-solving" mode, which characterizes what Thomas Kuhn called "normal science." What this complaint fails to acknowledge is that the word "science" applies equally to work at the edges of a paradigm in the confrontation with genuine anomalies, which involves the messy business of blazing a path in uncharted territory. Unfortunately, whether a problem represents a mere puzzle or a paradigm-busting anomaly is not usually clear except in retrospect.

(6) *Eyewitness testimony is unreliable.* The reliability of eyewitness testimony has been extensively studied by forensic psychologists (National Research Council, 2014; Loftus, 2019), including specifically in the case of UAP reports (Hynek, 1972a; Haines, 1980). It is noteworthy that the conclusions of this work are not that eyewitness accounts are completely without value or that they cannot provide useful and accurate information, but rather that the reports of witnesses and the witnesses themselves must be evaluated carefully in order to assess credibility (Hynek, 1972a, b; Haines, 1980). The astronomer J. Allen Hynek served as a consultant to UFO investigation efforts at the U.S. Air Force for two decades and confirmed, contrary to popular misconceptions, that (a) people who are not "UAP believers" do witness and report UAP; (b) reliable, stable, educated people report UAP; (c) scientifically trained people report UAP; and (d) some UAP reports include detailed descriptions of objects seen at close range.

For purposes of the present study, patterns abstracted from many eyewitness accounts, along with some instrument observations, are used to inform choices about instrumentation. This is not dissimilar to the way anecdotal evidence commonly motivates the design of experimental studies in medicine. The gold standard of scientific work is to make quantitative measurements using well-calibrated instruments under well-understood conditions, and this is the approach taken in this work.